\documentclass[orivec,conference]{llncs}
\usepackage[cmex10]{amsmath}
\usepackage{stmaryrd}
\usepackage{amsfonts}
\usepackage{amssymb}
\usepackage{bbold}
\usepackage{wrapfig}
\usepackage{graphicx}
\usepackage{enumerate}
\usepackage{multirow}
\usepackage{xspace}
\usepackage{cancel}
\usepackage{enumitem}
\usepackage[hidelinks]{hyperref}
\usepackage{esvect}
\usepackage{caption}
\usepackage{cancel}
\usepackage{xcolor,colortbl}
\newcommand{\tlp}{\mbox{$\mi{TRV}\!_{\mi{lp}}$}\xspace}
\newcommand{\rpu}{$\mi{Res}\hspace{-0.5pt}^{\mi{pu}}$\xspace}
\newcommand{\rbc}{$\mi{Res}\hspace{-0.5pt}^{\mi{bc}}$\xspace}
\newcommand{\rlp}{$\mi{Res}\hspace{-0.5pt}^{\mi{lp}}$\xspace}
\newcommand{\bcl}{\ti{BlkCls}\xspace}
\newcommand{\slp}{\ti{SimLP}\xspace}
\newcommand{\vlp}{\ti{VerLP}\xspace}
\newcommand{\flp}{\ti{FormLP}\xspace}
\newcommand{\lc}{\ti{lp}-clause\xspace}
\newcommand{\lcs}{\ti{lp}-clauses\xspace}
\newcommand{\elp}{\mbox{$\mi{EFF}\!_{\mi{lp}}$}\xspace}

\newcommand{\lp}{\mbox{$\mi{SAT}\!_{\mi{lp}}$}\xspace}

\newcommand{\ec}{$\mi{FM}\!_{\mi{ec}}$\xspace}
\newcommand{\tfm}{$\mi{TRV}\!_{\mi{fm}}$\xspace}
\newcommand{\fm}{\ti{FM}\xspace}
\newcommand{\kis}{\ti{Kissat}\xspace}
\newcommand{\mprop}{Proposition~\ref{prop:loc_prf}\xspace}

\newcommand{\vic}[1]{$#1$-vicinity\xspace}
\newcommand{\lpr}[1]{$#1$-certificate\xspace}
\newcommand{\lprs}[1]{$#1$-certificates\xspace}

\newcommand{\ppr}[1]{\mbox{$\mi{AppPr6}^*$}\xspace}

\newcommand{\slt}[2]{\mbox{$\mi{SatLits}(#1,\pnt{#2})$}\xspace}
\newcommand{\sslt}[2]{\mbox{$\mi{SatLits}(#1,\pent{#2}{^*})$}\xspace}

\newcommand{\lt}[1]{\mbox{$\mi{Lits}(#1)$}\xspace}

\newcommand{\egp}{\mbox{$\mi{EG}$-$\mi{PQE}^+$}\xspace}

\newcommand{\bm}[1]{{\mbox{\boldmath $#1$}}}
\newcommand{\Bm}[1]{{\boldmath $#1$}}

\newcommand{\imp}{\Rightarrow} 
\newcommand{\ie}{i.e.,\xspace}
\newcommand{\eg}{e.g.,\xspace}
\newcommand{\pnt}[1]{\mbox{$\vv{#1}$}\xspace}
\newcommand{\pnnt}[2]{\mbox{$\vv{#1}{#2}$}}

\newcommand{\peent}[3]{\mbox{$\vv{#1}#2_{#3}$}\xspace}
\newcommand{\pent}[2]{\mbox{$\vv{#1}#2$}}

\newcommand{\cof}[2]{\mbox{$#1_{\vec{#2}}$}}

\newcommand{\V}[1]{\mbox{$\mathit{Vars}(#1)$}}
\newcommand{\Va}[1]{\mbox{$\mi{Vars}(\vec{#1})$}}

\newcommand{\s}[1]{\mbox{$\{#1\}$}}

\newcommand{\nGz}[2]{$G_{non-\{z\}}$}

\newcommand{\prr}[1]{\mi{Prev}(\boldsymbol{q})}

\newcommand{\mi}[1]{\mathit{#1}}
\newcommand{\ti}[1]{\textit{#1}}
\newcommand{\tb}[1]{\textbf{#1}}

\newcommand{\ttt}{\>\>\>}
\newcommand{\tttt}{\>\>\>\>}
\newcommand{\ttttt}{\>\>\>\>\>}
\newcommand{\tttttt}{\>\>\>\>\>\>}

\newcommand{\Tt}{\>\>}
\newcommand{\Sub}[2]{\mbox{$\mi{#1}\!_{\mi{#2}}$}}

\newcommand{\prob}[2]{\mbox{$\exists{#1} [#2]$}}

\newcommand{\Comment}[1]{}

\begin{document}

\title{On Structure-Aware SAT Solving}


\author{Eugene Goldberg}
\institute{\email{eu.goldberg@gmail.com}}

\maketitle

\begin{abstract}
Although modern SAT solvers have made remarkable progress, they still
exploit formula structure only to a limited extent. We present two
techniques that make SAT solving more structure-aware and provide
experimental results showing their promise. The first technique
``$\mathit{melts}$'' the formula by reducing interaction between its
parts.  This is achieved by adding new variables, using existential
quantifiers, and applying a procedure called partial quantifier
elimination. The second technique exploits the notion of a
$\mathit{local}$ $\mathit{proof}$. The latter establishes that a formula is
unsatisfiable by proving that certain literals of its clauses are
redundant. We show that using local proofs makes SAT solving more
formula-specific.
\end{abstract}


%
\section{Introduction}
\label{sec:intro}
Given a Boolean formula\footnote{In this paper, we consider only
  formulas in conjunctive normal form (CNF).} $F(X)$, the SAT problem
is to determine if there is an assignment to the variables of $X$ for
which $F$ evaluates to 1. SAT solving plays a big role in practical
applications. Over the last three decades, SAT solvers have made
remarkable progress due to two major developments. First,
conflict-driven clause learning (\tb{CDCL}) was
introduced~\cite{grasp}, enabling solvers to prove unsatisfiability by
constructing an explicit resolution proof~\cite{resolution}. Second,
researchers developed effective decision-making tailored to
CDCL~\cite{chaff}. Nevertheless, general-purpose SAT solving faces
some challenges. First, building a practical SAT solver based on a
proof system more powerful than resolution is a tall order. Second,
even if a short resolution proof exists, a CDCL solver may struggle to
find it.

The problems above stem from non-automatizability. A propositional
proof system is \ti{automatizable} if a polynomial-time algorithm
exists to find a shortest proof in that
system~\cite{automat}. Unfortunately, non-trivial proof systems,
including resolution, are believed to be non-automatizable unless
some proposition like $P=\mi{NP}$ holds~\cite{res_nonautom}. Moreover,
increasing the power of a proof system entails a blow-up of the proof
space. Thus, the more powerful a proof system becomes, the less
amenable it may be to automation.

The issue of non-automatizability can be addressed by exploiting the
specifics of the formula at hand, \ie by making SAT solving more
structure-aware. In this paper, \tb{structure} refers to both a
formula's \ti{macrostructure}, such as cuts or semantic blocks of
clauses, and its \ti{microstructure}, such as the distribution of
literals among clauses. Thus, ``structure-aware'' is essentially
synonymous with ``formula-specific'', and even a random formula has
some structure.

Existing structure-aware algorithms typically employ CDCL with
heuristics tailored to a particular application (see
e.g.~\cite{circ_sat1,circ_sat2}).  In this paper, we introduce two
approaches to structure-aware SAT solving that differ substantially
from conventional CDCL methods.  We also give \ti{experimental}
evidence showing the promise of either approach. The first approach,
called \tb{formula melting} (\tb{\fm}), aims to make a formula
\ti{sparser} by introducing new variables, employing existential
quantifiers and applying a technique called \ti{partial quantifier
  elimination}~\cite{hvc-14}. The last one is a structure-aware
technique: its result depends on the specifics of the quantified
formula. So, this technique enables \fm to exploit the formula's
structure.
The second approach employs \ti{formula-specific} \tb{local
  proofs}. Namely, it exploits the fact that a formula can be proved
unsatisfiable by demonstrating the redundancy of certain literals in
some of its clauses.

The two approaches are mutually beneficial. For example, the
efficiency of \fm depends critically on the efficiency of partial
quantifier elimination. Arguably, the latter can be improved
substantially through the use of local proofs.  The rest of the paper
is organized as follows.  Section~\ref{sec:basic} gives the basic
definitions. Section~\ref{sec:backgr} provides some background on CDCL
SAT solving.  Section~\ref{sec:frm_mlt} presents \fm and
Section~\ref{sec:fm_for_sac} explains how to use \fm for
structure-aware computing. Sections~\ref{sec:fm_for_ec}
and~\ref{sec:exper_ec} discuss how to use \fm for equivalence checking
and present experimental results. Sections~\ref{sec:loc_prf}
and~\ref{sec:exper_lp} introduce local proofs and provide experimental
justification. Sections~\ref{sec:sat_lp} and~\ref{sec:bits} discuss
the use of local proofs in SAT solving. Finally,
Section~\ref{sec:dir_concl} presents our conclusions.

\section{Some Basics}
\label{sec:basic}

In this section, when we say ``formula'' without mentioning
quantifiers, we mean ``a quantifier-free formula''.
\subsection{Basic definitions}
\begin{definition}
\label{def:cnf}
We assume that formulas have only Boolean variables.  A \tb{literal}
of a variable $v$ is either $v$ (the positive literal) or its negation
$\overline{v}$ (the negative literal).  A \tb{clause} is a disjunction
of literals. A formula $F$ is in conjunctive normal form (\tb{CNF}) if
$F = C_1 \wedge \dots \wedge C_k$ where $C_1,\dots,C_k$ are
clauses. We will also view $F$ as the \tb{set of clauses}
\s{C_1,\dots,C_k}. We assume that \tb{every formula is in CNF} unless
otherwise stated.
\end{definition}

%
%
\begin{definition}
  \label{def:vars} Let $F$ be a formula. Then \bm{\V{F}} denotes the
set of variables of $F$, and \bm{\V{\prob{X}{F}}} denotes
$\V{F}\!\setminus\!X$.
\end{definition}

%
%
\begin{definition}
Let $V$ be a set of variables. An \tb{assignment} \pnt{q} to $V$ is a
mapping $V'~\rightarrow \s{0,1}$ where $V' \subseteq V$.  We will
denote the set of variables assigned in \pnt{q}~~as \bm{\Va{q}}. We will
refer to \pnt{q} as a \tb{full assignment} to $V$ if $\Va{q}=V$.
\end{definition}

%
%
\begin{definition}
A literal and a clause are \tb{satisfied} (respectively
\tb{falsified}) by an assignment \pnt{q} if they evaluate to 1
(respectively 0) under \pnt{q}.
\end{definition}

%
%
\begin{definition}
\label{def:cofactor}
Let $C$ be a clause. Let $H$ be a formula that may have quantifiers,
and \pnt{q} be an assignment to
\V{H}.  If $C$ is satisfied by \pnt{q}, then \bm{\cof{C}{q} \equiv
  1}. Otherwise, \bm{\cof{C}{q}} is the clause obtained from $C$ by
removing all literals falsified by \pnt{q}. Denote by \bm{\cof{H}{q}}
the formula obtained from $H$ by removing the clauses satisfied by
\pnt{q} and replacing every clause $C$ unsatisfied by \pnt{q} with
\cof{C}{q}.
\end{definition}

%
%
\begin{definition}
  \label{def:Xcls}
Given a formula \prob{X}{F}, a clause $C$ of $F$ is called
\tb{quantified} if \V{C} $\cap\, X \neq \emptyset$. Otherwise, the
clause $C$ is called \tb{unquantified}.
\end{definition}

%
%
\begin{definition}
\label{def:formula-equiv}
Let $G, H$ be formulas that may have existential quantifiers. We say
that $G, H$ are \tb{equivalent}, written \bm{G \equiv H}, if $\cof{G}{q} =
\cof{H}{q}$ for all full assignments \pnt{q} to $\V{G} \cup \V{H}$.
\end{definition}

%
%
\begin{definition}
\label{def:red_cls}
Let $F(X,Y)$ be a formula and $G \subseteq F$ and $G \neq
\emptyset$. The clauses of $G$ are said to be \textbf{redundant in}
\bm{\prob{X}{F}} if $\prob{X}{F} \equiv \prob{X}{F \setminus G}$. If
$F \setminus G$ implies $G$, the clauses of $G$ are redundant in
\prob{X}{F}, but the reverse is not true.
\end{definition}

%
%
\begin{definition}
 \label{def:pqe_prob} Given a formula \prob{X}{F(X,Y))} and $G$ where
 $G \subseteq F$, the \tb{Partial Quantifier Elimination} (\tb{PQE})
 problem is to find $H(Y)$ such that\linebreak \Bm{\prob{X}{F}\equiv
 H\wedge\prob{X}{F \setminus G}}.  (So, PQE takes $G$ out of the scope
 of quantifiers.)  The formula $H$ is called a \tb{solution} to
 PQE. The case of PQE where $G = F$ is called \tb{Quantifier
 Elimination} (\tb{QE}).
\end{definition}

%
%
\begin{example}
\label{exmp:pqe_exmp}
Consider the formula $F = C_1 \wedge \dots \wedge C_5$ where
$C_1=\overline{x}_3 \vee x_4$,\linebreak $C_2\!=\!y_1\!\vee\!x_3$,
$C_3=y_1 \vee \overline{x}_4$, $C_4\!=\!y_2\!\vee\!x_4$,
$C_5\!=\!y_2\!\vee \overline{x}_4$. Let $Y =\s{y_1,y_2}$ and $X =
\s{x_3,x_4}$. Consider the PQE problem of taking $C_1$ out of
\prob{X}{F} \ie finding $H(Y)$ such that $\prob{X}{F} \equiv H \wedge
\prob{X}{F \setminus \s{C_1}}$. One can show that $\prob{X}{F} \equiv
y_1 \wedge \prob{X}{F \setminus \s{C_1}}$, \ie $H\!  =\!y_1$ is a
solution to this PQE problem $\blacksquare$
\end{example}

%
%
\begin{proposition}
\label{prop:sol_for_pqe}
  Formula $H(Y)$ is a solution to the PQE problem of taking $G$ out of
  \prob{X}{F(X,Y)} (\ie $ \prob{X}{F} \equiv H \wedge \prob{X}{F
    \setminus G}$) iff \\ a) $F \imp H$ and b) $H \wedge \prob{X}{F}
  \equiv H \wedge \prob{X}{F \setminus G}$
\end{proposition}

(This proposition is proved in~\cite{cav23}.)  So, to take $G$ out of
\prob{X}{F(X,Y)}, it suffices to find $H(Y)$ implied by $F$ that makes
$G$ \ti{redundant} in $H \wedge \prob{X}{F}$.

Note that PQE is a \ti{structural} (\ie formula-specific) operation in
the following sense.  Assume $F \equiv F'$ and both $F$ and $F'$
contain a subset of clauses $G$.  The fact that $\prob{X}{F} \equiv H
\wedge \prob{X}{F \setminus G}$ \ti{does not} imply that $\prob{X}{F'}
\equiv H \wedge \prob{X}{F' \setminus G}$.  Thus, one cannot prove $H$
correct using the truth table of $F$ \ti{alone} because the
correctness of $H$ depends on the particulars of $F$.  That is, $H$ is
\ti{formula-specific}. (On the contrary, a solution $H$ to the QE
problem is \ti{not} formula-specific.  One can verify that
$\prob{X}{F} \equiv H$ using $H$ and the \ti{truth table} of $F$.)

%
%
\begin{definition}
\label{def:res_of_cls}
Let clauses $C'$, $C''$ have opposite literals of exactly one variable
$w\!\in\!\V{C'}\!\cap\!\V{C''}$.  Then $C'$, $C''$ are called
\tb{resolvable} on~$w$.  Let $C$ be the clause consisting of the
literals of $C'$ and $C''$ minus those of $w$. Then $C$ is said to be
obtained by \tb{resolution} of $C'$ and $C''$ on $w$.
\end{definition}

%
%
\begin{definition}
\label{def:blk_cls}
Let $C$ be a clause of a formula $F$ and $w \in \V{C}$. The clause $C$
is called \tb{blocked} in $F$ at $w$ if no clause of $F$ is resolvable
with $C$ on $w$~\cite{blocked_clause}.
\end{definition}
%
%
\refstepcounter{subsection}
\vspace{5pt}
\tb{\thesubsection\,\,Representing a circuit by a formula}
\label{ssec:circ_to_cnf}
\vspace{5pt}

\noindent Let $M(X,V,W)$ be a combinational circuit where $X,V$, and
$W$ are the sets of internal, input, and output variables of $M$,
respectively. Let $F(X,V,W)$ denote a formula specifying $M$. As
usual, this formula is obtained by Tseitsin's
transformations~\cite{tseitin}. Namely, $F=F_{g_1} \wedge \dots \wedge
F_{g_k}$ where $g_1,\dots,g_k$ are the gates of $M$ and $F_{g_i}$
specifies the functionality of gate $g_i$.
%
%
\begin{example}
\label{exmp:gate_cnf}
Let $g$ be a 2-input AND gate defined as $x_3\!=\!x_1\wedge\,x_2$ where
$x_3$ denotes the output variable and $x_1,x_2$ denote the input
variables. Then $g$ is specified by the formula
$F_g\!=\!(\overline{x}_1 \vee \overline{x}_2\vee x_3) \wedge (x_1 \vee
\overline{x}_3) \wedge (x_2 \vee \overline{x}_3)$. Every clause of
$F_g$ is falsified by an inconsistent assignment (where the output of
$g$ is not implied by its input). For instance, $x_1\!\vee
\overline{x}_3$ is falsified by the inconsistent assignment
$x_1\!=\!0, x_3\!=\!1$. So, every assignment \ti{satisfying} $F_g$
corresponds to a \ti{consistent} assignment to $g$ and vice
versa. Similarly, every assignment satisfying the formula $F$ above is
a consistent assignment to the gates of $M$ and vice versa
$\blacksquare$
\end{example}

\section{Some Background On CDCL SAT Solving}
\label{sec:backgr}
This section gives some background on CDCL SAT solving.  In
Subsection~\ref{ssec:prf_str} we argue that a reason for the success
of modern solvers is that they respect the general structure of a
resolution \ti{proof}. Subsections~\ref{ssec:macro_str} and
~\ref{ssec:micro_str} assert, however, that current SAT algorithms
can not fully exploit the structure of the \ti{formula}.
%
%
\subsection{Exploiting the structure of resolution proof}
\label{ssec:prf_str}
A resolution proof that a formula $F$ is unsatisfiable can be
represented as a directed acyclic graph $G$.  The sources of $G$
correspond to clauses of $F$, and its sink corresponds to an empty
clause. Every internal node of $G$ corresponds to a clause obtained by
resolving the clauses of the two incoming nodes.

Let a set of nodes of $G$ be a cut, and \Sub{F}{cut} be the set of
clauses corresponding to the nodes of this cut. Note that the nodes
between this cut and the sink form a resolution proof that
\Sub{F}{cut} is unsatisfiable~\cite{novikov}. So one can build $G$ by
constructing a sequence of unsatisfiable formulas \Sub{F}{cut}
starting with the cut defined by the sources of $G$ and ending with
the cut specified by the sink. Importantly, producing \Sub{F}{cut}
corresponding to a cut makes every clause $C$ assigned to a node
\ti{below} this cut \ti{redundant}. (Because $\Sub{F}{cut} \imp C$.)
Modern CDCL solvers implicitly use this property via decision-making
introduced by Chaff~\cite{chaff} that is biased towards recent
conflict clauses. This bias produces new conflict clauses that make
original clauses and earlier derived conflict clauses redundant. So,
the success of Chaff-like CDCL solvers
(e.g.~\cite{berkmin,minisat,picosat}) can be attributed to their
respect for the basic structure of a \ti{resolution proof}.
%
%

\subsection{Struggling with exploiting macrostructure}
\label{ssec:macro_str}
Assume $F$ has some macrostructure (\eg some non-trivial cuts). $F$
can have many resolution proofs of quite different sizes. So,
respecting the structure of a proof guarantees that it is built
\ti{reasonably} but not that it is necessarily
\ti{short}. Intuitively, since a short proof consists of a relatively
few resolutions, it should follow the structure of $F$. Let $F$, for
instance, specify equivalence checking of combinational circuits $M'$
and $M''$ that are structurally similar. In this case, a short
resolution proof of equivalence can be obtained by deriving clauses
relating internal points of $M'$ and $M''$ (see,
\eg~\cite{brand,kuehlmann97,fmcad16}). These relation clauses are
computed in topological order, moving \ti{cutwise} from inputs to
outputs of $M'$ and $M''$.  So, this proof of equivalence indeed
follows the macrostructure of $F$.

Unfortunately, CDCL solvers are not structure-aware enough to produce
the short proof above. The reason is that CDCL SAT solving is largely
semantic rather than structural in the following sense.  A CDCL solver
is essentially a search algorithm that examines the truth table of $F$
rather than follows its structure.  (Even though it performs this
examination in large chunks specified by conflict clauses.)  In
Sections~\ref{sec:frm_mlt}-\ref{sec:exper_ec}, we discuss a technique
called \ti{formula melting} that helps find short proofs via
exploiting the formula's structure.

\subsection{Struggling with exploiting microstructure}
\label{ssec:micro_str}
Another problem with CDCL SAT solving is that a resolution proof is
``unnecessarily strong''. Derivation of an empty clause from $F$ means
proving $F \imp 0$. This implication lets us conclude that $F$ is
unsatisfiable without knowing any \ti{specifics} of $F$. In
Sections~\ref{sec:loc_prf}\,-\ref{sec:bits}, we introduce the notion
of a \ti{local proof}. This proof shows that $F$ is unsatisfiable by
establishing that $F \imp G$, where $G$ is a \ti{satisfiable}
formula. This implication entails unsatisfiability only for a
\ti{particular} formula $F$. So, local proofs allow one to exploit
even the \ti{microstructure} of $F$.

\section{SAT Solving By Formula Melting}
\label{sec:frm_mlt}
One can address the CDCL problems above by designing SAT solvers that
exploit the formula's structure.  In
Sections~\ref{sec:frm_mlt}-\ref{sec:fm_for_ec}, we describe a
technique called \ti{f}\!ormula \ti{m}elting (\tb{\ti{FM}}). These
sections ignore the issue of the efficiency of \fm, which heavily
depends on the efficiency of PQE.  We discuss this issue in
Section~\ref{sec:exper_ec}.
\subsection{The basics of \fm}
Let $F(X)$ be a formula. \fm consists of a sequence of ``melting
steps''. The objective of each step is to produce a formula
equisatisfiable to $F$ that has a (slightly) different structure. If
\fm does not solve $F$, one can \tb{finish} the problem by applying a
regular SAT solver to the formula \fm produces.

Let $T$ be a subset of $X$. \tb{Melting} $F$ in variables
of $T$ consists of two substeps.  First, the set $T^*$ of new ``twin''
variables is added. For every variable $x \in T$, the set $T^*$
contains a twin variable $x^*$. Then, a formula $F^*$ is built from
$F$ by replacing variables of $T$ in \ti{some} clauses of $F$ with
their twins from $T^*$. Finally, a formula $\mi{Eq}(T,T^*)$ is formed
specifying the equivalence between variables of $T$ and their
corresponding twins in $T^*$. So, the formulas $F(X)$ and
$F^*(X,T^*)\wedge\mi{Eq}$ are equisatisfiable. We will refer to the
first substep as \tb{twinning} the variables of $T$. In the second
substep, $X \cup T^*$ is partitioned into subsets $X'$ and $X''$ of
quantified and unquantified variables, respectively. The set $T^*$ is
always in $X'$. Then, one performs PQE by taking $\mi{Eq}$ out of
\prob{X'}{\mi{Eq} \wedge F^*}, \ie by finding $H(X'')$ such that
$\prob{X'}{\mi{Eq} \wedge F^*} \equiv H \wedge \prob{X'}{F^*}$. Note
that the formula $H \wedge F^*$ is equisatisfiable to $F$.

The melting step above reduces the number of clauses resolvable on
variables of $X'$ at the expense of adding new clauses depending only
on $X''$. So, the clauses become more \ti{independent} of each other
in terms of $X'$. This is somewhat similar to turning a solid object
into liquid, \ti{hence the name} ``melting''. \fm can be viewed as a
special case of formula relaxation. In the example below, we consider
a step of \fm where a single new variable is added \ie $|T|=|T^*|=1$.
%
%
\begin{example}
\label{exmp:mlt}
Assume $F(X) = C_1 \wedge \dots \wedge C_4 \wedge \dots \wedge C_k$
where $C_1 = x_1 \vee x_2$, $C_2 = x_1 \vee x_3$, $C_3 =
\overline{x}_1 \vee x_4$, $C_4 = \overline{x}_1 \vee x_5$.  Assume
that $C_1,C_2,C_3,C_4$ are the only clauses of $F$ with variable
$x_1$. So, there are only four resolutions on $x_1$: $C_1$ with
$C_3$,$C_4$, and $C_2$ with $C_3,C_4$. Below we describe melting $F$
in $T = \s{x_1}$.

In the first substep, the new twin variable $x^*_1$ is added, \ie
$T^*\!=\!\s{x^*_1}$. Assume that one modifies $F$ into a formula $F^*$
by replacing the variable $x_1$ in $C_1$ and $C_3$ with the new
variable $x^*_1$. That is $C_1$ is modified into $C^*_1 = x^*_1 \vee
x_2$ and $C_3$ into $C^*_3 = \overline{x}^*_1 \vee x_4$. Finally, the
formula $\mi{Eq}(T,T^*)$ is built where
$\mi{Eq}=\mi{eq}(x_1,x^*_1)$. Here, $\mi{eq}$ specifies the
equivalence of $x_1$ and $x^*_1$ and consists of clauses $x_1 \vee
\overline{x}^*_1$ and $\overline{x}_1 \vee x^*_1$. Note that $\mi{Eq}
\wedge F^*$ is equisatisfiable to $F$ since the former is just $F$
where $x_1$ is replaced with an equivalent variable $x^*_1$ in some
clauses.

In the second substep, one picks a subset $X'$ of $X \cup \s{x^*_1}$
as the set of quantified variables in formula \prob{X'}{\mi{eq} \wedge
  F^*} where $x^*_1 \in X'$. (Assume that $x_1$ is too in $X'$
although this is optional.)  One concludes the second substep by
taking out $\mi{eq}$, \ie by building $H(X'')$ such that
\prob{X'}{\mi{eq} \wedge F^*} $\equiv H \wedge \prob{X'}{F^*}$. Here
$X''$ is the set of unquantified variables \ie $X'' = (X \cup
\s{x^*_1}) \setminus X'$.  Note that due to the lack of $\mi{eq}$ in
$H \wedge F^*$, the variable $x^*_1$ of the clauses $C^*_1,C^*_3$ is
now \ti{independent} of $x_1$ of $C_2,C_4$.  We will say that
$C^*_1,C^*_3$ and $C_2,C_4$ \tb{get separated} by melting $F$ in
$x_1$.
So, instead of four possible resolutions on $x_1$ in $F$, the formula
$H \wedge F^*$ has only two possible resolutions on $x_1$ and
$x^*_1$. Indeed, $C^*_1$ can still be resolved with $C^*_3$ on $x^*_1$,
and $C_2$ can still be resolved with $C_4$ on $x_1$. But $C^*_1$
cannot be resolved with $C_4$ and $C_2$ cannot be resolved with
$C^*_3$ $\blacksquare$
\end{example}

\subsection{Some discussion of \fm}
\label{ssec:fm_compl}
%
%
One can use multiple steps of melting in single variables to build a
SAT procedure. (It is described in more detail in
Appendix~\ref{app:str_fm}.) This is a trivial procedure and so it is
not particularly practical.  We mention it here to underscore \fm's
completeness. Assume, for instance, that the set of unquantified
variables $X''$ is the same \ti{small} subset of $X$ for all melting
steps. (On the other hand, the set of quantified variables $X'$ grows
due to adding twin variables.)  Each subsequent melting step leads to
reducing the number of possible resolutions on variables of $X'$ and
adds new clauses depending only on $X''$. Let \Sub{F}{X''} be the
subset of all clauses of $F$ depending only on $X''$. Multiple melting
steps have two outcomes.  If $F$ is unsatisfiable, \Sub{F}{X''}
eventually becomes unsatisfiable. Otherwise, all resolutions on
variables of $X'$ eventually go away without making \Sub{F}{X''}
unsatisfiable. So, to check the satisfiability of $F$, it suffices to
check that of \Sub{F}{X''}. Assuming that $X''$ is small, \Sub{F}{X''}
can be solved by a trivial SAT solver based on DPLL~\cite{dpll}.

%

The procedure above is somewhat similar to performing QE on
\prob{X'}{F} by elimination of quantified variables. The difference is
that instead of eliminating $x\in X'$, a step of \fm adds a twin
variable $x^*$ to $x$ and makes both weaker than the original $x$.
Let $F_x$ denote the subset of clauses of $F$ having a literal of $x$.
The specifics of twinning $x$ depend on the \ti{subset} of clauses of
$F_x$ in which $x$ is replaced with $x^*$. (For instance, in
Example~\ref{exmp:mlt}, one could replace $x_1$ with $x^*_1$ in
clauses $C_2,C_4$ instead of $C_1,C_3$.)  So, the total number of
different ways to replace $x$ with $x^*$ is \ti{exponential} in
$|F_x|$.

\section{Using \fm For Structure-Aware Computations}
\label{sec:fm_for_sac}
As mentioned above, a step in \fm has exponentially many
versions. Here we explain how one can use this freedom to perform
\ti{structure-aware} computations.

\subsection{A simple example}
Let $F$ be a formula equal to $A(X,Y) \wedge B(Y,Z)$, where $X,Y,Z$ are
non-overlapping sets of variables. Suppose one wants to produce a
simpler formula equisatisfiable to $F$ via exploiting the structure of
$F$, \ie the existence of the cut specified by $Y$. A naive approach is
to build a formula $I(Y)$ equivalent to \prob{X}{A} by QE. Then $F$
and $I \wedge B$ are equisatisfiable. But such $I$ can be
prohibitively large.

Below we show how $F$ can be simplified by melting. Consider melting
$F$ in variables of $Y= \s{y_1,\dots,y_k}$.  Denote by $Y^*$ the set
of twin variables, \ie $Y^* = \s{y^*_1,\dots,y^*_k}$, where $y^*_i$ is
the twin variable of $y_i$, $i=1,\dots,k$. Let $\mi{Eq}(Y,Y^*)$ denote
the formula $\mi{eq}(y_1,y^*_1) \wedge \dots \wedge
\mi{eq}(y_k,y^*_k)$ specifying the equivalence of variables of $Y$ and
$Y^*$. Assume a formula $F^*$ is produced by replacing the variables
of $Y$ with twin ones \ti{only} in the clauses of $A$. So, $F^* =
A(X,Y^*) \wedge B(Y,Z)$. Note that $\mi{Eq} \wedge F^*$ is
equisatisfiable to $F$.

Assume that, in the second substep of melting, $\mi{Eq}$ is taken out
of\, \prob{W}{\mi{Eq} \wedge F^*} where $W = X \cup Y^* \cup Z$, \ie
only $Y$ remains unquantified. Let $H(Y)$ be a solution to the PQE
problem above, \ie \prob{W}{\mi{Eq} \wedge F^*}~$\equiv$ $H \wedge
\prob{W}{F^*}$.  Assuming that $A$ is satisfiable, $H \wedge
\prob{W}{F^*} \equiv H \wedge \prob{W}{B}$. Then $F$ is
\ti{equisatisfiable} to $H(Y) \wedge B(Y,Z)$.  And the latter is a
correct simplification of $F$.  If $A \wedge B$ is unsatisfiable and $A
\imp H$, then $H$ is an \tb{interpolant}~\cite{craig,ken03}. If $A
\wedge B$ is satisfiable, the satisfying assignments of $H \wedge B$
can be viewed as an \tb{abstraction} of those of $F$. Namely, every
assignment (\pnt{y},\pnt{z}) satisfying $H \wedge B$ can be extended
to an assignment (\pnt{x},\pnt{y},\pnt{z}) satisfying $A \wedge B$.
(\pnt{x},\pnt{y},\pnt{z} are full assignments to $X,Y,Z$.)

\subsection{Some discussion}
Note that the choices made when forming \prob{W}{\mi{Eq} \wedge F^*}
above, heavily exploited the existence of the cut $Y$ \ie the
\tb{structure} of $F$. Indeed, in general, any \ti{subset of
  variables} of $F$ could be picked for twinning instead of $Y$. And
even if $Y$ is chosen, \ti{any subset of clauses} of $F$ with
variables of $Y$ could be used for replacement with twin ones (instead
of using only clauses of $A$). Finally, when taking $\mi{Eq}$ out of a
quantified version of $\mi{Eq} \wedge F^*$ one could pick any subset
of \V{F^*} containing $Y^*$ as quantified variables (rather than $W =
X \cup Y^* \cup Z$).

The choices above were meant to \tb{separate} $A$ and $B$. The fact
that adding $H$ enables this separation means that $H$ ``absorbs'' the
structure of $F$. Note that $H$ can be much smaller than a formula
obtained by QE on \prob{X}{A}. Let \pnt{x},\pnt{y},\pnt{z} be full
assignments to $X,Y,Z$, respectively.  The presence of a clause
falsified by \pnt{y} in $H$ is mandatory \ti{only} if \pnt{y} can be
extended to (\pnt{y},\pnt{z}) satisfying $B$ but not to
(\pnt{x},\pnt{y}) satisfying $A$. So, in contrast to QE, $H$ requires
only the presence of clauses describing the \ti{interaction} of
formulas $A$ and $B$.

The idea of separation extends to the case where $F$ describes
\tb{many} interacting blocks. Full separation is just a special case
of \fm. For instance, one could melt the formula $A \wedge B$ in a
\ti{subset} of $Y$ just to \ti{reduce} the interaction of $A$ and $B$.
Note that when a \tb{CDCL} solver applies to the formula $A \wedge B$,
it actually \tb{increases} the interaction of $A$ and $B$ by adding
clauses that relate variables of $X$ and $Z$. While this
structure-unaware strategy still works well for many practical
formulas, it may fail for some applications (e.g. equivalence
checking).

\section{Using \fm For Equivalence Checking}
\label{sec:fm_for_ec}
In this section, we apply \fm to equivalence checking of combinational
circuits. Earlier an equivalence checking algorithm based on PQE was
presented~\cite{fmcad16}. We show that this algorithm is a
\ti{special} case of \fm (a \ti{general} method). This justifies
using equivalence checking to compare \fm and CDCL.

\subsection{Equivalence checking by \fm}
\label{ssec:fm_ec}
Let $M'(X,Y',z')$ and $M''(X,Y'',z'')$ be single-output combinational
circuits. Here, $X$ is the set of input variables and $Y'$ and $Y''$
are the sets of internal variables of $M'$ and $M''$, respectively.  The
variables $z'$ and $z''$ specify the outputs of $M'$ and $M''$,
respectively. Circuits $M',M''$ are called \ti{equivalent} if they
produce identical values of $z',z''$ for every input \pnt{x} (\ie a
full assignment to $X$).

Let $F'(X,Y',z')$ and $F''(X,Y'',z'')$ specify $M'$ and $M''$
respectively as described in Subsection~\ref{ssec:circ_to_cnf}.  Let
$F$ denote the formula $\overline{\mi{eq}} \wedge F' \wedge F''$,
where $\mi{eq}(z',z'')$ specifies the equivalence of $z'$ and
$z''$. The circuits $M'$ and $M''$ are \tb{functionally equivalent}
iff the formula $F$ is unsatisfiable

Below we describe a method of equivalence checking by \fm. We will
refer to this method as {\boldmath \ec}.  This method is based on
\tb{melting} $F$ in the variables of $X$. Let $X=\s{x_1,\dots,x_n}$
and $X^*$ denote the set of twin variables \s{x^*_1,\dots,x^*_n}, where
$x^*_i$ is the twin of $x_i$, $i=1,\dots,n$. Let $F^*$ denote the
formula $F$ in which the variables of $X$ are replaced with
corresponding twins \ti{only} in the clauses of $F''$.  So $F^*=
\overline{\mi{eq}}(z',z'') \wedge F'(X,Y',z') \wedge
F''(X^*,Y'',z'')$. Let $\mi{Eq}(X,X^*)$ specify the equivalence of
variables of $X$ and $X^*$, \ie $\mi{Eq} = \mi{eq}(x_1,x^*_1) \wedge
\dots \wedge \mi{eq}(x_n,x^*_n)$. Then $F$ is equisatisfiable to
$\mi{Eq}(X,X^*) \wedge F^*$.

Assume that in the second substep of \fm, the variables of $W = X \cup
X^* \cup Y' \cup Y''$ are chosen to be \tb{quantified}. (So, only the
variables $z',z''$ are unquantified.) Let $H(z',z'')$ be the formula
obtained by taking $\mi{Eq}$ out of $\prob{W}{\mi{Eq} \wedge
  F^*}$. That is $\prob{W}{\mi{Eq} \wedge F^*}$ $\equiv H \wedge
\prob{W}{F^*}$.  The proposition below (proved in~\cite{fmcad16})
substantiates \ec.

\begin{proposition}
\label{prop:ec_by_pqe}
Assume $M',M''$ do not implement a constant (0 or 1). Then $M'$ and
$M''$ are equivalent iff $H(z',z'') \imp \mi{eq}(z',z'')$.
\end{proposition}

\vspace{-3pt}

If $H \imp \mi{eq}(z',z'')$, then $H \wedge
\overline{\mi{eq}}(z',z'')$ is unsatisfiable and hence $H \wedge F^*$
and $F$ are unsatisfiable too. So equivalence checking of $M'$ and
$M''$ \ti{comes down} to melting the formula $F$ in variables of
$X$. (Checking that $M'$ and $M''$ are not constants reduces to a few
easy SAT checks.)
%
%
\subsection{Cutwise computation}
\begin{wrapfigure}{l}{1.9in}
 \begin{center}
    \includegraphics[width=1.8in]{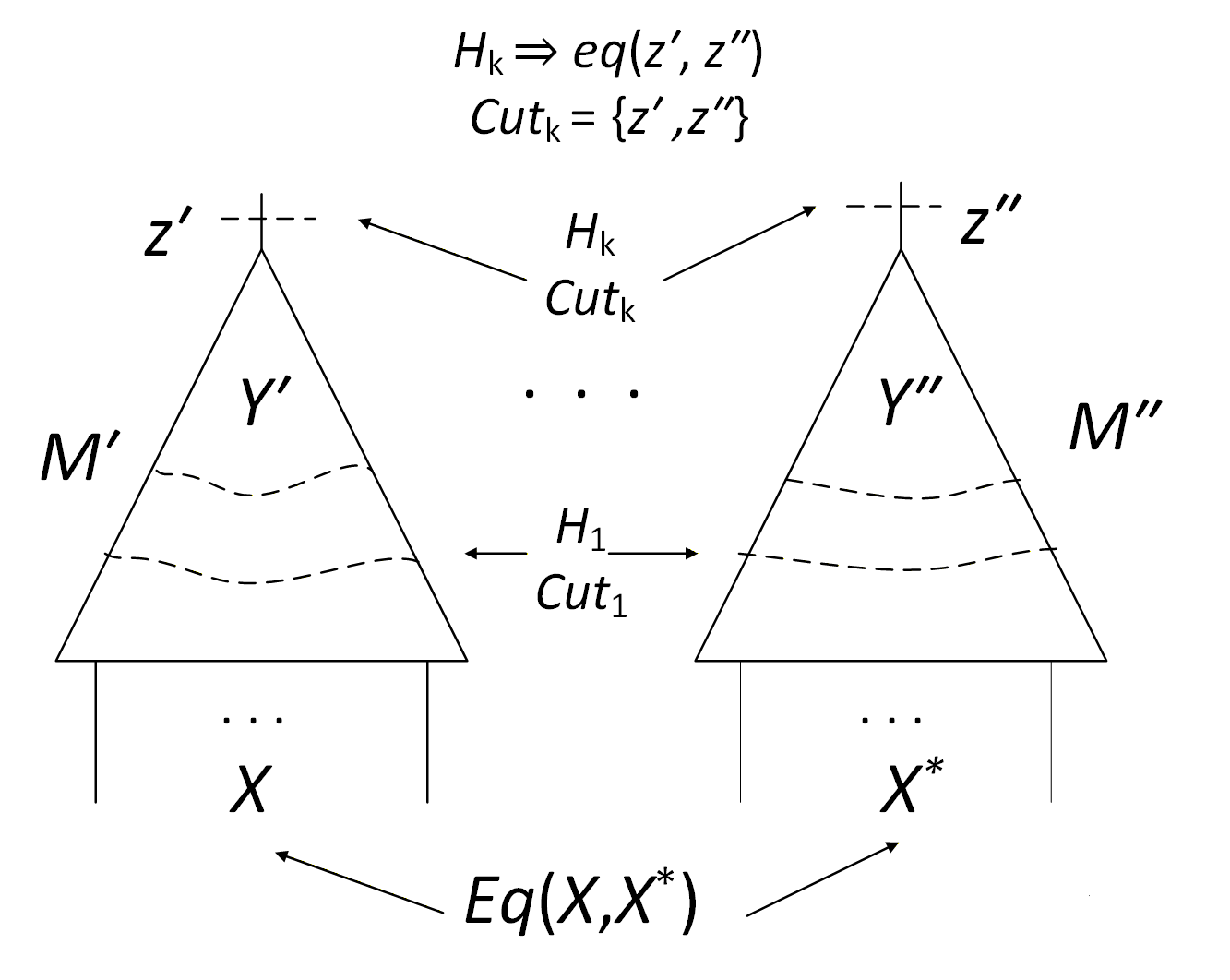}
  \end{center}
\vspace{-10pt}
\caption{Cutwise computation}
\vspace{3pt}
\label{fig:cutwise}
\end{wrapfigure}

To make taking out $\mi{Eq}$ easier, \ec does it incrementally. Let
$\mi{Cut}_1,\dots,\mi{Cut}_k$ be a set of cuts shown in
Fig.~\ref{fig:cutwise}, where $\mi{Cut}_k = \s{z',z''}$.  The formula
$H$ above is computed \tb{cutwise}. Namely, one uses PQE to build a
sequence of formulas $H_1,\dots,H_k$, each formula $H_i$ depending
only on variables of $\mi{Cut}_i$. The last formula $H_k$ specifies
$H$ of Proposition~\ref{prop:ec_by_pqe}.  Below, we describe a
simplified computation of formulas $H_i$. (A more efficient method
that was used in the experiment of Section~\ref{sec:exper_ec} is
described in Appendix~\ref{app:fm_ec}.)

\setlength{\intextsep}{4pt}

Let $W_i$ specify the variables of $W$ minus those of
$\mi{Cut}_i$. Since $W$ does not contain the variables of $\mi{Cut}_k$
(\ie $z',z''$), then $W_k = W$.  Let $H_1$ be a formula obtained by
taking $\mi{Eq}(X,X^*)$ out of \prob{W_1}{\mi{Eq} \wedge F^*}, \ie
\prob{W_1}{\mi{Eq} \wedge F^*} $\equiv H_1 \wedge
\prob{W_1}{F^*}$. Let $H_i$, $i=2,\dots,k$ denote the formula obtained
by taking $H_{i-1}$ out of \prob{W_i}{H_{i-1} \wedge F^*}, \ie
\prob{W_i}{H_{i-1} \wedge F^*} $\equiv H_i \wedge
\prob{W_i}{F^*}$. The \tb{semantics} of the sequence $H_1,\dots,H_k$
is quite simple. Every next formula $H_i$ makes the previous formula
$H_{i-1}$ \tb{redundant} (because $H_{i-1}$ is redundant in $H_i
\wedge \prob{W_i}{H_{i-1} \wedge F^*}$). So, $\mi{Eq}$ and
$H_1,\dots,H_{i-1}$ are redundant by the time $H_i$ is produced.  This
essentially means that $M'$ and $M''$ are \tb{separated} below
$\mi{Cut}_i$.

One can show that each formula $H_i$ is also a solution to the PQE
problem of taking $\mi{Eq}$ out of \prob{W_i}{\mi{Eq} \wedge F^*}, \ie
\prob{W_i}{\mi{Eq} \wedge F^*} $\equiv H_i \wedge \prob{W_i}{F^*}$,
$i=1,\dots,k$. Since $W_k = W$, the formula \bm{H_k} \tb{specifies}
the formula $H$ of Proposition~\ref{prop:ec_by_pqe}.

\subsection{Some discussion}
\label{ssec:ec_dis}
Melting the formula $F$ in variables of $X$ essentially separates
circuits $M'$ and $M''$. So, the formula $H$ obtained by \fm that
implies their equivalence, in a sense, ``absorbs'' the macrostructure
of $F$. The only exception occurs when $M'$ and $M''$ are identical
constants. In this case, $M'$ and $M''$ are equivalent even if they
are separate. (So, $\mi{Eq}$ is simply redundant in $\prob{W}{\mi{Eq}
  \wedge F^*}$.)

\setlength{\intextsep}{4pt}
\begin{wrapfigure}{l}{1.8in}
 \begin{center}
    \includegraphics[width=1.7in]{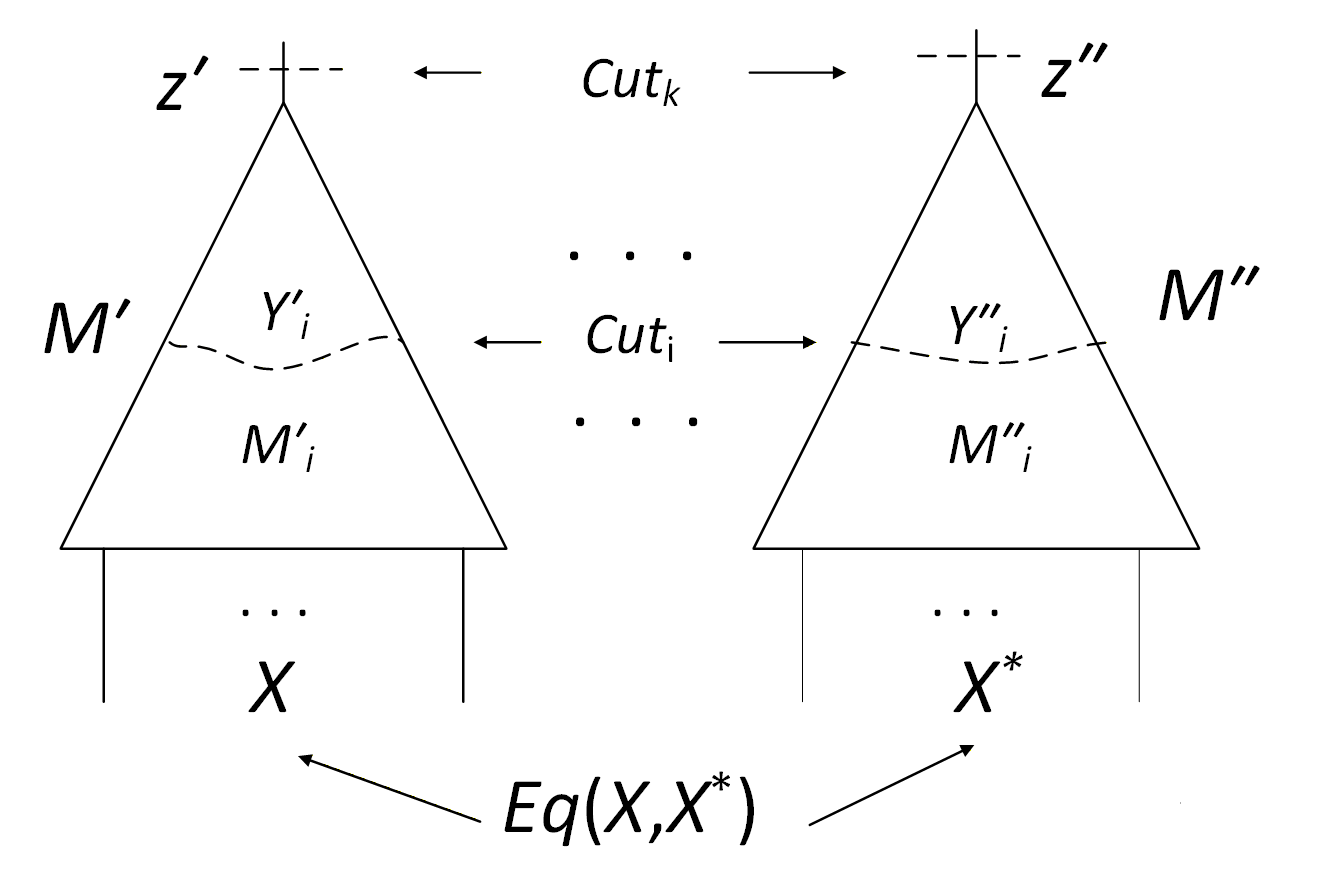}
  \end{center}
\vspace{-10pt}
\caption{A cut}
\label{fig:sngl_cut}
\end{wrapfigure}


Below, we show that if $M'$ and $M''$ are structurally similar, \ec
can generate a short proof unlike a CDCL SAT solver.  Consider the
sequence of formulas $H_1,\dots,H_k$. As mentioned above, $H_i$ can be
viewed as a solution to the PQE problem of taking $\mi{Eq}$ out of
\prob{W_i}{\mi{Eq} \wedge F^*}. Let $M'_i$ and $M''_i$ specify the
gates of $M'$ and $M''$ respectively that are located below
$\mi{Cut}_i$ (see Fig.~\ref{fig:sngl_cut}). Let $Y'_i$ and $Y''_i$
specify the parts of $\mi{Cut}_i$ corresponding to the output
variables of $M'_i$ and $M''_i$ respectively.  So, $\mi{Cut}_i = Y'_i
\cup Y''_i$.

A full assignment to $\mi{Cut}_i$ can be represented as
(\peent{y}{'}{i},\peent{y}{''}{i}), where
\peent{y}{'}{i},\peent{y}{''}{i} are full assignments to $Y'_i$ and
$Y''_i$, respectively. Note that $H$ \ti{does not} have to \smallskip
contain a clause falsified by an assignment
(\peent{y}{'}{i},\peent{y}{''}{i}) if, say, \peent{y}{'}{i} cannot be
produced by $M'_i$ \ti{individually}. Such a clause is implied by
$F^*$ that remains in the formula after taking out $\mi{Eq}$.  The
formula $H$ \ti{has to} contain a clause falsified by
(\peent{y}{'}{i},\peent{y}{''}{i}) \ti{only} if \peent{y}{'}{i} and
\peent{y}{''}{i} can be produced by $M'_i$ and $M''_i$ individually
but \ti{cannot} be produced \ti{together} due to the constraint
$\mi{Eq}(X,X^*)$. So, $H$ must only exclude assignments to
$\mi{Cut}_i$ that cannot appear because of the \ti{interaction}
between $M'$ and $M''$, \ie because of the \tb{structure} of $\mi{Eq}
\wedge F^*$. It was shown in~\cite{fmcad16} that $H_i$ is small if
$M'$ and $M''$ are structurally similar. In particular, if the value
of every cutpoint $y \in Y'_i$ can be expressed as a function of the
values of no more than $m$ cut points of $Y''_i$, then $H_i$ can be
composed of clauses with no more than $m+1$ literals. So, if $m$ is a
constant, $H_i$ has a polynomially bounded size.


Generating a short proof by \ec (when $M'$ and $M''$ are similar)
relies on gradually separating $M'$ and $M''$. A CDCL SAT solver
cannot produce such a proof for the following reason. The separation
above results from replacing $\mi{Eq}$ with $H_1$ and every $H_i$ with
$H_{i+1}$, where $0 < i < k$. This replacement is based on the notion
of redundancy in a \ti{quantified formula} (see
Definition~\ref{def:red_cls}). So, a CDCL solver cannot do such
replacement.  In reality, this solver does the \tb{opposite} of
separation by adding clauses that relate the internal variables of
$M'$ and $M''$ and, thus, increasing the interaction of $M'$ and
$M''$.


\section{An Experiment With Equivalence Checking}
\label{sec:exper_ec}
The viability of \fm crucially depends on the efficiency of PQE
solving, which is still work in progress.  In this section, we conduct
an experiment showing that current PQE solvers are good enough to make
\fm viable for at least one important problem: equivalence checking.
(All experiments of this paper were run on a computer with Intel Core
i5-10500 CPU @ 3.1GHz.)
In our experiment, we apply \ec to check the equivalence of circuits
$M'$ and $M''$ that are deliberately constructed to make the problem
hard for both specialized equivalence checkers and CDCL SAT solvers.
Either circuit is an $m\text{-}$bit multiplier with an additional
Boolean input $w$.  If $w=1$, both $M'$ and $M''$ work as
$m\text{-}$bit multipliers. If $w=0$, both $M'$ and $M''$ output the
number 0. So, $M'$ and $M''$ are equivalent. The difference is in
their structure.

%
\begin{wraptable}{l}{2.2in}
\small
\caption{\ti{Equivalence checking of $M'$ and $M''$ built off a multiplier}}
\vspace{-10pt}
\setlength\extrarowheight{3pt}
\scriptsize
\begin{center}
\begin{tabular}{|c|c|c|c|c|} \hline
 \#bits  &  \multicolumn{2}{c|}{\kis} &    \multicolumn{2}{c|}{\ec}  \\ \cline{2-5}
           & \#cnfls  & time     & \#cnfls  &    time          \\ \hline
 10   & $6.5\!\times\!10^4$  & 1.4s &$1.4\!\times\!10^4$ & 2.0s\\ \hline
 11  & $2.4\!\times\!10^5$ & 5.8s& $1.8\!\times\!10^4$ & 3.0s\\ \hline
12  & $3.8\!\times\!10^5$& 8.3s& $2.3\!\times\!10^4$ &4.4s\\ \hline
13  &  $5.9\!\times\!10^5$&16s &$2.9\!\times\!10^4$ &6.1s \\ \hline
14  &$1.1\!\times\!10^6$&39s &$3.6\!\times\!10^4$&  9.0s\\ \hline
15   & $1.3\!\times\!10^6$ &  41s &  $4.4\!\times\!10^4$ &  12s \\ \hline
16    &  $2.4\!\times\!10^6$  &1m\,40s&  $5.1\!\times\!10^4$  & 16s       \\ \hline\hline
   24    &   $3.8\!\times\!10^7$  & 58m\,20s & $1.7\!\times\!10^5$ & 1m\,51s              \\ \hline
   32     & $2.5\!\times\!10^8$  & 17h\,3m    &   $3.8\!\times\!10^5$  & 8m\,21s    \\ \hline
   40     &  $1.1\!\times\!10^9$  & 155h\,13m  &  $7.3\!\times\!10^5$   &  30m\,51s        \\ \hline
\end{tabular}                
\end{center}
\vspace{-10pt}
\label{tbl:ec_vs_kis}
\end{wraptable}

In $M'$ every \ti{input} variable of one operand is conjoined with
$w$. So, when $w=1$ it works like a regular multiplier, but when
$w=0$, this operand specifies the number 0 and $M'$ outputs 0. In
$M''$, every \ti{output} variable is conjoined with $w$. So, when
$w\!=\!1$, $M''$ operates like a regular multiplier, but when
$w\!=\!0$, it outputs 0.  Note that $M'$ and $M''$ do not have
functionally \ti{equivalent} internal \ti{points}.  (The output of
every internal gate of $M'$ depends on $w$, in contrast to that of
$M''$.)  This makes equivalence checking of $M'$ and $M''$ hard even
for specialized algorithms that look for the points above.


\ec is a special case of \fm, a \tb{general} technique applicable to
\ti{any} CNF formula. So, we compare \ec with \kis~\cite{kissat}, a
high-quality CDCL SAT solver.  In particular, we compare the size of
proofs generated by \ec and \kis. When implementing \ec, we used \egp,
the latest PQE solver~\cite{cav23} downloaded
from~\cite{eg_pqe_plus}. In the experiment, we used \ec and \kis to
check whether the middle bits of $M'$ and $M''$ where
equivalent. Namely, we applied \ec and \kis to the formula $F$
described in Subsection~\ref{ssec:fm_ec}.

Table~\ref{tbl:ec_vs_kis} shows the experimental results.  The first
column gives the size of the multiplier used to form $M'$ and
$M''$. The next two columns give the results of \kis (number of
conflicts and run time.) The last two columns show \ec results. \egp
employs branching and forms a solution from conflict clauses. The
first column of the \ec results gives the \ti{total} number of
conflicts accrued across \ti{all} calls of \egp.
Table~\ref{tbl:ec_vs_kis} shows that \ec effectively exploited the
structure of $F$. It solved the formula \tb{faster} because it
produced a much \tb{shorter} proof. Subsection~\ref{ssec:ec_dis}
explains why. Assuming an average conflict clause has the same number
of resolutions for \ec and \kis, the proof generated by \ec for the
40-bit $M'$ and $M''$ was at least 1000 times shorter. (\kis actually
used many more resolutions than \ec per conflict clause.)


\vspace{-1pt}
\section{Proving Unsatisfiability Locally}
\label{sec:loc_prf}
This section starts the second part of this paper
(Sections~\ref{sec:loc_prf}-\ref{sec:bits}) where we describe a method
of structure-aware SAT solving based on the notion of a local
proof. Importantly, even if a formula has no \ti{macrostructure}, a
local proof can still exploit its \ti{microstructure}.  In
Subsection~\ref{ssec:clust} we introduce the key notions of a cluster
of clauses and a certificate clause.  Subsection~\ref{ssec:loc_prf}
presents local proofs.


%
%
  \subsection{Some definitions}
\label{ssec:clust}
%
%
\begin{definition}
  \label{def:prim_sec} Let $C$ be a clause of formula
$F$. Let \bm{F_c} be the set of all clauses of $F$ having at least one
literal of $C$ (including $C$ itself). We will refer to $F_c$ as
a \tb{cluster} of clauses of $F$ \tb{specified} by $C$. We will call
$C$ the \tb{primary} clause of $F_c$ and the clauses of
$F_c \setminus \s{C}$ the \tb{secondary} clauses of $F_c$.
\end{definition}
%
%
\begin{definition}
  \label{def:intern_lit} Let $C$ be a clause of $F$. Let $B$ be a
clause of the cluster $F_c$ specified by $C$ and $l$ be a literal of
$B$.  We will call $l$ a \tb{primary} literal of $F_c$ if it is also a
literal of the primary clause $C$.
\end{definition}
%
%
\begin{example}
Let $F$ denote $C_1 \wedge \dots \wedge C_4 \wedge \dots \wedge C_k$
where $C_1\!=\!x_1\!\vee\!x_2$,\linebreak $C_2 =
x_1 \vee \overline{x}_7 \vee x_9$, $C_3 = x_1 \vee \overline{x}_3$,
$C_4 = x_2 \vee x_5 \vee x_6$. Assume $C_1,\dots,C_4$ are the only
clauses of $F$ with literals $x_1$ and $x_2$. (But no restriction is
imposed on clauses of $F$ with $\overline{x}_1$ and/or
$\overline{x}_2$.)  $C_1$ specifies the cluster $F_{c_1}$ equal
to \s{C_1,C_2,C_3,C_4}. Here $C_1$ is the primary clause, and
$C_2,C_3,C_4$ are the secondary clauses of $F_{c_1}$.  Literals $x_1$
and $x_2$ are the primary literals of $F_{c_1}$ $\blacksquare$
\end{example}


%
%
\begin{definition}
\label{def:red_lit}
Let $C$ be a clause of a formula $F$ and $l$ be a literal of $C$. The
literal $l$ is called \tb{redundant} in $C$ if removing it from $C$
does not change the truth table of $F$. 
\end{definition}

%
%
\begin{definition}
\label{def:vicin}
Let $C$ be a clause of $F$ and $l$ be a literal of $C$.  The
assignment satisfying $l$ and falsifying the other literals of $C$ is
called the \bm{l}\tb{-vicinity} of $C$.
\end{definition}
%
%
\begin{proposition}
\label{prop:red_lit}
A literal $l$ of a clause $C$ of a formula $F$ is redundant in $C$ iff
the subspace specified by the \vic{l} of $C$ has no assignment
satisfying $F$.
\end{proposition}

The proofs of all new propositions are given in
Appendix~\ref{app:proofs}.  In this paper, we will use the phrase
``literal $l$ of $C$ is redundant'' as a \tb{shorter version} of ``the
\vic{l} of $C$ has no satisfying assignment''. That is, establishing
redundancy of $l$ is not meant for ``optimizing'' the formula by
removing this literal from $C$.

%
%
\begin{definition}
\label{def:cert}
Let $l$ be a literal of a clause $C$ of a formula $F$. A clause $B$ is
called an \bm{l}\tb{-certificate} for $C$ if $F \imp B$ and $B$ is
falsified by the \vic{l} of $C$.
\end{definition}

According to Proposition~\ref{prop:red_lit}, the clause $B$ above
certifies that $l$ is redundant in $C$, hence the name ``\lpr{l}''.

%
%
\begin{example}
  \label{exmp:vic}
Let $C\!=\!x_1 \vee x_2 \vee x_3$ be a clause of $F$. Then, say, the
$x_1$-vicinity of $C$ is specified by the assignment
\pnt{q}\!=($\bm{x_1\!=\!1},x_2\!=\!0, x_3\!=\!0$). (The assignment to
$x_1$ satisfying $C$ is shown in bold.)  $B\!=\!\overline{x}_1\!\vee
x_2\!\vee\!x_3$ is an \lpr{x_1} clause for $C$ if $F\!\imp\!B$. Then
no assignment of subspace \pnt{q} satisfies $F$ and the literal $x_1$
is redundant. Clauses $x_2\!\vee\!x_3$ and $\overline{x}_1$ implied by
$F$ are also \lprs{x_1} for $C$ $\blacksquare$
\end{example}

\subsection{Local proofs}
\label{ssec:loc_prf}
%
%
The proposition below improves the result of~\cite{provinglocally}.
\begin{proposition}
\label{prop:lit_vic} Let $C$ be a clause of a formula $F$. Let $F_c$
be the cluster of $F$ specified by $C$.  If $F$ is satisfiable, there
exists a satisfying assignment \pnt{x}, a clause $B \in F_c$ and a
primary literal $l$ of $B$ such that \pnt{x} is in the $l$-vicinity of
$B$. (So, according to Proposition~\ref{prop:red_lit}, the literal $l$
is \ti{not redundant} in $B$.)
\end{proposition}
%
%
\begin{proposition}
  \label{prop:loc_prf}
Let $C$ be a clause of a formula $F$. Let $F_c$ be the cluster of $F$
specified by $C$. Let $P$ be a formula such that for every clause $B
\in F_c$ and for every primary literal of $B$, $P$ contains an
\lpr{l}.  Then $F$ is unsatisfiable.
\end{proposition}

We will refer to $P$ as a \tb{local proof} of unsatisfiability of $F$
because it is based on examining only a \ti{part} of the search
space. (This part consists of $l$-vicinities explored to derive
$l$-certificates.) For that reason, $P$, in general, is a satisfiable
formula. So, a local proof is \tb{formula-specific} because the fact
that $F$ implies a satisfiable formula $P$ can entail unsatisfiability
only for a \ti{particular} formula $F$. This differs from, say, a
resolution proof that asserts $F \imp 0$. The latter claim implies
that $F$ is unsatisfiable \ti{regardless} of its specifics. Note that
our definition of a local proof \ti{abstracts away} the details of
deriving $P$. (A full proof of unsatisfiability of $F$ obviously
requires providing those details.)

\begin{example}
\label{exmp:prop}
Here is an example of a local proof.  Let $F=C_1 \wedge C_2 \wedge C_3
\wedge\dots\wedge C_k$ where $C_1 = x_1 \vee \overline{x}_2$, $C_2 = x_1 \vee
x_5$, $C_3 = \overline{x}_2 \vee \overline{x}_6 \vee x_8$.  Assume
$C_1,C_2,C_3$ are the only clauses of $F$ with literals $x_1$ and/or
$\overline{x}_2$, whereas $F$ can contain any number of clauses with
literals $\overline{x}_1$ and/or $x_2$.  Consider the cluster
$F_{c_1}=\s{C_1,C_2,C_3}$, where $C_1$ is the primary clause and
$C_2,C_3$ are secondary clauses.
According to \mprop, to prove $F$ unsatisfiable, it suffices to show
that the literals $x_1$ and $\overline{x}_2$ are redundant in $C_1$,
and that the literals $x_1$ and $\overline{x}_2$ are redundant in
$C_2$ and $C_3$, respectively.  That is, there is no assignment
satisfying $F$ in subspaces $(\bm{x_1\!=\!1},x_2\!=\!1)$,
$(x_1\!=\!0,\bm{x_2\!=\!0})$ and $(\bm{x_1\!=\!1},x_5\!=\!0)$ and
$(\bm{x_2\!=\!0},x_6\!=\!1,x_8\!=\!0)$.

Assume $F \imp P$ where $P = B'_1 \wedge B''_1 \wedge B_2 \wedge B_3$
and $B'_1 = \overline{x}_1 \vee \overline{x}_2$, $B''_1 = x_1
\vee x_2$, $B_2 = \overline{x}_1 \vee x_5$, $B_3 = x_2 \vee
\overline{x}_6 \vee x_8$. Here $B'_1$ and $B''_1$ are $x_1$-certificate and
$\overline{x}_2$-certificate for $C_1$, and $B_2$ and $B_3$ are
$x_1$-certificate and $\overline{x}_2$-certificate for $C_2,C_3$,
respectively. According to \mprop, the set $P$ forms a \tb{local
proof} of unsatisfiability of $P$. Note that $P$ itself is
satisfiable. It remains satisfiable even when conjoined with
$F_{c_1}$. So, one cannot derive an empty clause from $F_{c_1} \wedge
P$ by resolution and needs to use \ti{the rest} of $F$.

The assignments satisfying $F_{c_1}\!\wedge\!P$ are
($x_1\!=\!1,x_2\!=\!0,x_5\!=\!1,x_6\!=\!0$) and
($x_1\!=\!1,x_2\!=\!0,x_5\!=\!1,x_8\!=\!1$). These assignments are
eliminated by adding a single \tb{blocked} clause $K
= \overline{x}_1 \vee x_2 \vee
\overline{x}_5$. (It is blocked in $F$ at $x_1$ since it is
unresolvable with $C_1$ and $C_2$ on $x_1$.) So, $F_{c_1} \wedge P
\wedge K$ is unsatisfiable $\blacksquare$
\end{example}

%
%

If $F_c \wedge P$ is satisfiable, proving $F$ unsatisfiable by
resolution \ti{alone} may still require a lot of work even if the
local proof $P$ is already available.  On the other hand,
Example~\ref{exmp:prop} implies that one can make $F_c \wedge P$
unsatisfiable by adding blocked clauses.  This can be viewed as
simulation of \mprop via the machinery of blocked clauses.  Resolution
+ blocked clauses is exponentially more powerful than resolution
alone~\cite{gen_ext_resol}. This substantiates the claim that
resolution + local proofs is likely more powerful than resolution
alone. (Appendix~\ref{app:blk_cls} describes experiments on simulation
of local proofs by blocked clauses. It provides data supporting the
claim that blocked clauses cannot efficiently simulate local proofs.)
We mention blocked clauses here to show how one
can \ti{simulate} \mprop.  \ti{Using} this proposition \ti{does not}
require generation of blocked clauses.

\section{Some Experimental Justification Of Local Proofs}
\label{sec:exper_lp}
In this section, we conduct the following experiment showing the
promise of local proofs as a \ti{formula-specific} technique. We start
with a formula $F$ and apply \kis, used earlier in
Section~\ref{sec:exper_ec}, to prove $F$ unsatisfiable.  Then we use a
simple procedure to build a set of clauses $P$ that forms a local
proof of unsatisfiability under the assumption $F \imp P$ (the details
are in Appendix~\ref{app:gen_loc_prf}).  On one hand, assuming $F \imp
P$, one can verify that $P$ is a local proof in almost linear
time. Namely, one needs to find a clause $C \in F$ such that $P$
contains all necessary certificate clauses for the cluster $F_c$ and
so \mprop holds. We will refer to this procedure as \tb{\vlp} where
\ti{Ver} stands for \ti{verify}.  (Appendix~\ref{app:lp_ver} gives its
pseudocode).  On the other hand, one can exploit the same assumption
$F\!\imp\!P$ via running \kis on $F \wedge P$. In this way, we measure
the distance from the local proof $P$ to a resolution proof $F \wedge
P\!\imp\!0$ that \kis has to cover.

\begin{wraptable}{l}{3.3in}
\small
\caption{\ti{Hard random formulas. Using $F\imp P$ \\by \kis and \vlp
    (via \mprop)}}
\label{tbl:rnd_form_res}
\vspace{-10pt}
\setlength\extrarowheight{3pt}
\scriptsize
\begin{center}
  \begin{tabular}{|c|c|c|c|c|c|c|c|c|} \hline
    name &  \multicolumn{2}{c|}{size of $F$}  &  \multicolumn{2}{c|}{solv. $F$, \kis} &$|P|$ &\multicolumn{3}{c|}{using $F\!\imp\!P$}\\ \cline{2-5} \cline{7-9}
    & \#vars    & \#cls   & \#\ti{cnfls.}&\ti{time} & & \multicolumn{2}{c|}{solv.\,$F \wedge P$,\,\kis} & \vlp\\ \cline{7-8}
    & & &   & & &\#\ti{cnfls.}&\ti{time} & \ti{time}\\ \hline
    $\mi{rnd}_1$ & 250 & 1,062   &  $1.8\!\ast\!10^5$ & 3.7s &11 & $8.1\!\ast\!10^4$  &   1.5s   & 0.002s \\ \hline
    $\mi{rnd}_2$ & 300 & 1,275  & $1.4\!\ast\!10^6$ &  1m\,6s &12 &  $4.7\!\ast\!10^5$ &  13s & 0.003s  \\ \hline
    $\mi{rnd}_3$ & 350 & 1,487  & $1.4\!\ast\!10^7$ & 18m &17 & $1.9\!\ast\!10^6$ & 1m\,6s     &  0.003s      \\ \hline
    $\mi{rnd}_4$ & 375 & 1,593  & $3.5\!\ast\!10^7$  & 1h\,54m &22 & $7.4\!\ast\!10^6$& 7m\,24s & 0.003s \\ \hline
     $\mi{rnd}_5$ & 400 & 1,700   & $4.8\!\ast\!10^7$   & 3h\,18m  &14 & $1.7\!\ast\!10^7$ & 23m &  0.004s \\ \hline
    $\mi{rnd}_6$ & 425 &  1,806  & $1.7\!\ast\!10^8$   & 24h & 19 & $4.0\!\ast\!10^7$ &  2h\,13m & 0.003s   \\ \hline
    $\mi{rnd}_7$ & 450 &  1,912 &  $\!6.6\!\ast\!10^8$   & 305h &16 & $\!1.9\!\ast\!10^8$ & 41h & 0.003s \\ \hline
  \end{tabular}
\end{center}
\vspace{-10pt}
\end{wraptable}

In this section, we use random 3-CNF formulas where the number of
clauses is $4.25$ times that of variables.  Such formulas are easy to
generate and hard to solve~\cite{rand_cnf}. We also use bounded model
checking (BMC) formulas~\cite{bmc}.

Table~\ref{tbl:rnd_form_res} shows the results for a sample of random
formulas. The first three columns give the name, number of variables,
and number clauses of the formula. The next two columns show the
performance of \kis when solving $F$ (the number of conflicts and
runtime in hours and/or minutes and/or seconds). The following column
gives the number of certificate clauses in $P$. The next two columns
describe the performance of \kis when solving the formula $F \wedge
P$. The last column shows the runtime of \vlp that as written in
\ti{Python}~\cite{python}.

The run time of \vlp under the assumption $F \imp P$ was very small,
whereas \kis needed extra time to prove $F \wedge P$ unsatisfiable,
\ie using the \ti{same} assumption as \vlp. (For $\mi{rand}_7$, that
extra time was \tb{41 hours}.) Our comparison is obviously
\ti{incomplete} since we ignore the complexity of proving $F \imp P$.
But even this incomplete comparison shows the significant promise of
local proofs. Our experiment also substantiates the claim that local
proofs can exploit even the \ti{microstructure} of the formula (random
formulas lack any \ti{macrostructure}).

\begin{wraptable}{l}{3.3in}
\vspace{3pt}
\small
\caption{\ti{BMC formulas. Using $F\imp P$ by \kis and \vlp (via
    \mprop) }}
\label{tbl:pc_form_res}
\vspace{-10pt}
\setlength\extrarowheight{3pt}
\scriptsize
\begin{center}
  \begin{tabular}{|c|c|c|c|c|c|c|c|c|} \hline
    name &  \multicolumn{2}{c|}{circuit size}  &  \multicolumn{2}{c|}{solv. $F$, \kis} &$|P|$ &\multicolumn{3}{c|}{using $F\!\imp\!P$}\\ \cline{2-5} \cline{7-9}
    &\#tf  &\#gates  & \#\ti{cnfls.}&\ti{time} & & \multicolumn{2}{c|}{solv. $F \wedge P$, \kis} & \vlp\\ \cline{7-8}
   & & & & & &\#\ti{cnfls.} &\ti{time} & \ti{time} \\ \hline
6s101& 100 & $6.2\!\ast\!10^5$     &  $9.8\!\ast\!10^5$ &3m  & 7&   $1.1\!\ast\!10^6$  & 4m\,30s       & 4.7s\\ \hline
6s154 & 4 &   $5.2\!\ast\!10^4$        &  $2.1\!\ast\!10^7$ & 26m &71&  $5.9\!\ast\!10^6$ & 5m\,42s        & 1.0s\\ \hline
6s165 & 10 &$1.5\!\ast\!10^4$     &   $7.2\!\ast\!10^6$ & 5m\,16s & 6&  $4.4\!\ast\!10^6$& 3m\,11s         & 0.1s\\ \hline
6s168 & 10 & $2.8\!\ast\!10^4$   &    $7.2\!\ast\!10^6$ & 3m\,25s &74&  $4.9\!\ast\!10^6$ & 2m\,46s         &0.4s \\ \hline
6s284 &100& $2.2\!\ast\!10^6$   &  $2.2\!\ast\!10^6$ & 16m  & 7&  $2.9\!\ast\!10^6$ & 15m        &14.8s \\ \hline
6s343 &20 &$9.6\!\ast\!10^5$    &  $2.2\!\ast\!10^6$ & 7m\,16s  &7 &  $2.7\!\ast\!10^6$& 5m\,43s          & 7.7s\\ \hline
6s372  & 100 &$8.3\!\ast\!10^5$   &  $7.2\!\ast\!10^6$ &34m  & 7&   $8.2\!\ast\!10^6$ & 32m            & 7.3s\\ \hline
\end{tabular}
\end{center}
\vspace{-10pt}
\end{wraptable}

Table~\ref{tbl:pc_form_res} provides results of the same experiment
for a sample of BMC formulas. The first column gives the benchmark
name from a HardWare Model Checking Competition (HWM-CC-13). The
second and third columns show the number of time frames and the number
of gates in the unfolded circuit. The other columns have the same
meaning as in Table~\ref{tbl:rnd_form_res}. The results here are
similar to those for random formulas. Despite the existence of a local
proof $P$, \kis had to do a lot of work to generate a \ti{resolution}
proof that $F \wedge P$ was unsatisfiable.

\section{A Trivial SAT Procedure Based On Local Proofs}
\label{sec:sat_lp}
%
%
\setlength{\intextsep}{2pt}
\setlength{\textfloatsep}{10pt}
\begin{wrapfigure}{L}{1.5in}
\small
\begin{tabbing}
aaa\=b\=cc\= dd\= \kill
$\tlp(F,\pnt{q},\pnnt{q}{'})$\\
\tb{\scriptsize{1}}\>  $\pnt{q}\!:=\!\mi{BCP}(F,\pnt{q},\pnnt{q}{'})$ \\
\tb{\scriptsize{2}}\> if ($\mi{FalsifCls}(F,\pnt{q})$) \\
\tb{\scriptsize{3}}\Tt return(\ti{false}) \\
\tb{\scriptsize{4}}\> $C:= \mi{PickUnsCls}(F,\pnt{q})$   \\
\tb{\scriptsize{5}}\>  if ($C = \mi{nil}$) return($\mi{true}$)\\
\tb{\scriptsize{6}}\> $F_c\!:=\!\mi{UnsClust}(F,C,\pnt{q})$ \\
~~~~$\text{------------------------------}$ \\
\tb{\scriptsize{7}}\> foreach $B \in F_c$   \\
\tb{\scriptsize{8}}\Tt foreach $l \in \mi{Lits}(B,\pnt{q})$\{\\
\tb{\scriptsize{9}}\ttt  if ($l \not\in \mi{Lits}(C)$) \\
\tb{\scriptsize{10}}\tttt continue \\
\tb{\scriptsize{11}}\ttt $\pnnt{q}{'}\!:=\!\mi{Vicin}(F,B,l,\pnt{q})$  \\
\tb{\scriptsize{12}}\ttt if ($\tlp\!(F,\pnt{q},\pnnt{q}{'})$)   \\
\tb{\scriptsize{13}}\tttt  return($\mi{true}$)\} \\
\tb{\scriptsize{14}}\> return($\mi{false}$)  \\
\end{tabbing} 
\vspace{-20pt}
\caption{\tlp}
\vspace{5pt}
\label{fig:triv_lp}
\end{wrapfigure}

 We will refer to the general technique of
using local proofs for SAT solving as \lp.  In this section, we
describe a trivial implementation of \lp called \tlp. We present \tlp
to show that a SAT procedure based on local proofs a) can be
\ti{complete} and b) can prove unsatisfiability \ti{without} examining
the entire search space. In a sense, \tlp plays the role of the DPLL
procedure~\cite{dpll} for \lp.

Fig.~\ref{fig:triv_lp} shows the pseudocode for \tlp. It checks the
redundancy of primary literals for a cluster of clauses chosen in
the current subspace. \tlp performs those checks by recursively
calling itself. It accepts the formula $F$ to solve, the current
partial assignment \pnt{q} and an assignment \pnnt{q}{'} to be added
to \pnt{q}. \tlp returns \ti{true} (respectively \ti{false}) if $F$ is
satisfiable (respectively unsatisfiable) in the subspace $\pnt{q} \cup
\pnnt{q}{'}$.

\tlp starts by running Boolean Constraint Propagation (BCP) to extend
the current assignment \pnt{q} to $\pnt{q} \cup \pnnt{q}{'}$ and
satisfy all unit clauses that appear.  (A clause is called unit if one
of its literals is unassigned and all others are falsified.) If a
clause gets falsified during BCP, \tlp returns \ti{false} (lines 2-3).
Otherwise, \tlp picks a clause $C$ that is not satisfied by \pnt{q}
(line 4). If all clauses of $F$ are satisfied by \pnt{q}, then \tlp
returns \ti{true} (line 5).  Otherwise, \tlp builds the cluster $F_c$
specified by $C$ (line 6). The clauses sharing a literal with $C$ but
satisfied by \pnt{q} are omitted from $F_c$.

In the second part (lines 7-14), \tlp checks if \mprop holds for $F_c$
without generating certificates \ti{explicitly}. For each clause $B
\in F_c$ (line 7) and each unassigned literal $l$ of $B$ shared with
$C$ (lines 8-10), \tlp does the following. First, it forms \pnnt{q}{'}
that is the \vic{l} of $B$ in subspace \pnt{q} (line 11). Namely,
\pnnt{q}{'} satisfies $l$ and falsifies every other unassigned literal
of $B$. Then \tlp recursively calls itself (line 12). If this call of
\tlp returns \ti{false}, $l$ is redundant in $B$ in subspace \pnt{q}
and the inner loop continues. (This means that an $l$-certificate of
redundancy \ti{exists} in subspace \pnt{q}.)  Otherwise, $F$ is
satisfiable in subspace \pnt{q} and \tlp returns \ti{true} (line
13). If \tlp reaches line 14, then \mprop holds for $F_c$ in subspace
\pnt{q} and so $F$ is unsatisfiable in this subspace.

\tlp is correct and complete due to \mprop. But unlike algorithms
using variable splitting, \tlp does \ti{not} explore the \ti{entire}
search space.

\section{On Efficient Implementation Of \lp}
\label{sec:bits}
Building an \ti{efficient} implementation of \lp is an unsolved
problem yet.  (We will refer to such an implementation as \bm{\elp}.)
We assume that \elp includes checking literal redundancy, running BCP,
identifying conflicts and local proofs, and learning clauses. In this
section, we describe some additional potential features of \elp:
learning so-called \lcs (Subsection~\ref{ssec:lp_cls}) and keeping
clusters of clauses small (Subsections~\ref{ssec:separ}
and~\ref{ssec:opt_clust}). A natural question is whether looking for
\elp is worth the effort. In Subsection~\ref{ssec:dep_lp}, we argue
that the machinery of local proofs allows \elp to employ
``two-directional'' learning that is unavailable for CDCL solvers and
that can considerably boost SAT solving.

%
%

%
%
\vspace{-4pt}
\subsection{Generation of \lcs}
\label{ssec:lp_cls}
Let $F$ be a formula to solve. Besides conflict clauses, \elp can
learn \bm{lp}\tb{-clauses}. An \lc is built if a local proof is found
in subspace \pnt{q}. It is falsified by \pnt{q} and implied by $F$.
(Appendix~\ref{app:lp_cls} gives an example of an \lc and a proof of
correctness of the way \lcs are built.)  An \lc contains only
literals \ti{relevant} to the existence of the local proof.  This is
similar to a conflict clause that contains only literals relevant to
the conflict. An \lc is also similar to a conflict clause in that it
asserts that $F$ is unsatisfiable in subspace \pnt{q}.

\vspace{-5pt}
\subsection{Keeping clusters small by separating formula and proof}
\label{ssec:separ}
Adding new clauses to $F$ makes clusters larger and finding local
proofs harder. This problem can be addressed in \elp by keeping the
learned clauses in a different set $P$, \ie by \tb{separating} the
formula $F$ and the ``proof'' $P$.  That is, a cluster is formed only
from clauses of $F$ rather than $F \wedge P$ and so, literal
redundancy is tested only for clauses of $F$. This ensures that
learned clauses do not change the size of clusters.  (Assume, for the
sake of simplicity, that every clause learned by \elp, \eg a conflict
clause, is a \ti{certificate}, asserting redundancy of a literal in
some subspace. Then $P$ is indeed a set of ``proof clauses''.)
Ignoring the clauses of $P$ is logically correct because $F \imp P$
and hence $P$ is \ti{redundant} in $F \wedge P$. So, one can check
\mprop for $F$ instead of $F \wedge P$. Importantly, the formula $F
\wedge P$ can \ti{still} be used for \ti{BCP} and conflict clause
generation: there is no reason to ignore $P$ here. Thus, learning new
\ti{conflict clauses} in \elp can be as powerful as in CDCL SAT
solving.

\subsection{Keeping clusters small via cluster optimization}
\label{ssec:opt_clust}
Now, assume that \elp maintains a less rigorous strategy than above.
Namely, it still separates the formula $F$ and the set of certificate
clauses $P$ but it adds some learned clauses to $F$. In this case,
cluster sizes can grow. Let $F_c$ be the cluster specified by a clause
$C \in F$. The proposition below shows that one can build a local
proof by constructing certificates only for a \tb{subset} $F'_c$ of
$F_c$. Informally, $F'_c$ is a minimal subset of $F_c$ such that one
can drop the clauses of $F_c \setminus F'_c$ from $F$ and still derive
the required certificates of literal redundancy for $F'_c$.  We will
refer to $F'_c$ as an \tb{optimized} cluster.
%
%
\begin{proposition}
  \label{prop:opt_loc_prf}
Let $C$ be a clause of $F$ and $F_c$ be the cluster of $F$ specified
by $C$.  Let $F'_c$ be a subset of $F_c$ that includes $C$ itself. Let
$F''_c=F_c \setminus F'_c$. Let $P$ be a set of certificate clauses
such that for every clause $B \in F'_c$ and for every primary literal
$l$ of $B$, the formula $P$ contains an \lpr{l} implied by $F
\setminus F''_c$.  Then $F\setminus F''_c$ (and hence $F$) is
unsatisfiable.
\end{proposition}

Proposition~\ref{prop:opt_loc_prf} shows that, using optimized
clusters, \elp can build \tb{small local proofs} even if $F$ grows
\ti{large}. But this requires extra effort to keep track of clauses
involved in the generation of certificates.

%

%
%
\subsection{Using  two-directional learning}
\label{ssec:dep_lp}
The separation of the formula $F$ and the proof $P$ mentioned above
has \ti{one more} justification: the requirements for clauses of $F$
and $P$ are \ti{different}. Namely, the clauses of $F$ are preferred
to be \tb{longer}, and the clauses of $P$ to be \tb{shorter}.  Proving
redundancy of a literal $l$ of a longer clause $C\in F$ is \ti{easier}
because the search subspace \pnt{q} to explore is smaller (\pnt{q}
falsifies all literals of $C$ but $l$). Yet, a shorter clause $B \in
P$ is more likely to be an \lpr{l} in the current subspace. So, \elp
can potentially benefit from learning \ti{both short and long}
clauses.

The features of \elp above strongly differentiate it from CDCL
solvers.  First, a CDCL solver is essentially a search algorithm in
which every clause specifies an unsatisfiable part of the search
space. So, there is no reason to separate formula and proof
clauses. Second, a CDCL solver uses a \tb{one-directional} strategy:
learn shorter and shorter clauses until an empty clause is
produced. So, deriving many long clauses \ti{hampers} its performance
because long clauses are \ti{further away} from the global goal, an
empty clause.

From a high-level point of view, \elp can do the following to amplify
the power of local proofs.  First, long clauses are added to $F$ and
short ones to $P$. Second, the heuristic above is combined with
cluster optimization. This enables building clusters with a heavy
presence of long clauses even if the original formula consisted of
short clauses. And clusters of long clauses make finding a local proof
much more efficient.
Using \tb{two-directional} learning in \elp (\ti{long} clauses for $F$
and \ti{short} for $P$) creates a \ti{completely new} direction for
research that can lead to a significant boost in SAT solving.

\section{Conclusions}
\label{sec:dir_concl}
We introduce two general methods of structure-aware SAT solving that
apply to \ti{any} formula. The method called \tb{formula melting}
(\fm) is based on adding new variables, using quantifiers, and
applying partial quantifier elimination (PQE). Our conclusions here
are as follows.
First, \fm can be very effective for SAT solving formulas with
\ti{macrostructure} because it can exploit the presence of cuts,
semantic blocks of clauses, and other related structures.  Second,
since \fm is based on PQE, building new powerful PQE solvers is
important for structure-aware computing. Third, even existing PQE
solvers are powerful enough to be practical for equivalence checking,
an important \fm application.

The second method is based on the notion of a \tb{local proof} that
establishes a formula's unsatisfiability by checking literal
redundancy. One can make the following conclusions. First, since local
proofs are formula-specific, they can exploit even the formula's
\ti{microstructure}. Second, resolution + local proofs is likely more
powerful than resolution alone. We substantiate this claim
\ti{experimentally}.  Third, a SAT solver employing local proofs can
potentially benefit from learning both long and short clauses.  In
summary, finding an efficient implementation of a SAT solver using
local proofs is worth the effort.

\bibliographystyle{IEEEtran}
\bibliography{short_sat,local}

\clearpage
\appendix
\noindent{\large \tb{Appendix}}

%
%
%

%
%

\section{Proofs}
\setcounter{proposition}{2}
\label{app:proofs}
\subsection{Propositions of Section~\ref{sec:loc_prf}}
%
%
\begin{proposition}
\label{prop:red_lit}
A literal $l$ of a clause $C$ of a formula $F$ is redundant in $C$ iff
the subspace specified by the \vic{l} of $C$ has no assignment
satisfying $F$.
\end{proposition}
\noindent\tb{The ``if'' part.} Assume that the subspace specified
by the \vic{l} of $C$ has no satisfying assignment. Then the clause
$C^*$ obtained from $C$ by flipping the literal $l$  is implied
by $F$. By resolving $C$ and $C^*$ one produces the clause
$C$ minus the literal $l$ that is implied by $F$.

\noindent\tb{The ``only if'' part.} Assume that $l$ is redundant in
$C$.  So the clause obtained by removing $l$ from $C$ is implied by
$F$. Hence $F$ cannot be satisfied by an assignment that falsifies all
the literals of the clause $C$ but $l$.

%
%
\begin{proposition}
  \label{prop:lit_vic}
Let $C$ be a clause of a formula $F$. Let $F_c$ be the cluster of $F$
specified by $C$.  If $F$ is satisfiable, there exists a satisfying
assignment \pnt{x}, a clause $B \in F_c$ and a primary literal $l$ of
$B$ such that \pnt{x} is in the $l$-vicinity of $B$. (So,
Proposition~\ref{prop:red_lit} entails that $l$ is \ti{not redundant}
in $B$.)
\end{proposition}

\begin{proof}
Let $X$ denote \V{F}. Let \pnt{p} be a full assignment to $X$
satisfying $F$.  Let \lt{C} denote the set of literals of $C$, \ie the
primary literals of $F_c$. Assume there is a clause $B \in F_c$ for
which \pnt{p} satisfies only one literal $l$ and $l \in \lt{C}$. Then
\pnt{p} is the satisfying assignment \pnt{x} we look for.
Now, assume that there is no clause $B$ above. Then \pnt{p} satisfies
at least two literals of the primary clause $C$.  Denote by \slt{C}{p}
the literals of $C$ satisfied by \pnt{p}.  (So, $\slt{C}{p} \subseteq
\lt{C}$.)  Below, we show that by flipping values of \pnt{p}
satisfying literals of \slt{C}{p}, one eventually obtains a satisfying
assignment \pnt{x} we look for.

Let $l(x)$ be a literal of \slt{C}{p} where $x \in X$. Denote by $Q$
the subset of $F_c$ having at least one literal from \slt{C}{p}. Note
that by our assumption, every clause of $Q$ has at least two literals
satisfied by \pnt{p}. Denote by \pent{p}{^*} the assignment obtained
from \pnt{p} by flipping the value of $x$. Let $B$ be a clause of $F$.
Note that if $B \not\in Q$, the number of literals satisfied by
\pent{p}{^*} is either the same as for \pnt{p} or increases by 1. (The
latter may occur if $B$ contains the literal $\overline{l(x)}$.)  Only
if $B \in Q$, the number of literals satisfied by \pent{p}{^*} may
decrease in comparison to \pnt{p}.  In particular, the number of
literals of the primary clause $C$ satisfied by \pent{p}{^*} is
decreased by 1.

Denote by \sslt{C}{p} the set of literals of $C$ satisfied by
\pent{p}{^*} (that is equal to $\slt{C}{p} \setminus \s{l(x)}$).  If
there is a clause of $Q$ for which \pent{p}{^*} satisfies only one
literal, then it is a satisfying assignment \pnt{x} we look for. If
not, we produce the set $Q^*$ obtained from $Q$ by removing the
clauses that do not contain a literal of \sslt{C}{p} and continue the
procedure above. Note that $Q^*$ and \pent{p}{^*} preserve the
property of $Q$ and \pnt{p}. That is \pent{p}{^*} satisfies at least
two literals of every clause of $Q^*$. Also, every clause removed from
$Q^*$ remains satisfied no matter how we flip the literals of
\sslt{C}{p}. (Because, every such a clause had at least two literals
satisfied by \pnt{p} and hence at least one literal satisfied by
\pent{p}{^*}.  And this literal will remain satisfied because it is
not \sslt{C}{p}.)

Now we pick a new literal $l(x^*)$ from \sslt{C}{p} and flip the value
of $x^*$. Eventually such a procedure will produce an assignment that
satisfies only one literal of either the clause $C$ itself or some
other clause of the current version of the original set $Q$.
\end{proof}

%
%
\begin{proposition}
  \label{prop:loc_prf}
Let $C$ be a clause of a formula $F$. Let $F_c$ be the cluster of $F$
specified by $C$. Let $P$ be a formula such that for every clause $B
\in F_c$ and for every primary literal of $B$, $P$ contains an
\lpr{l}.  Then $F$ is unsatisfiable.
\end{proposition}

\begin{proof}
The formula $F$ cannot have any of the satisfying assignments
(described in Proposition~\ref{prop:lit_vic}) that every satisfiable
formula is supposed to have. (Each such an assignment falsifies some
certificate of $P$.) Hence, $F$ is unsatisfiable.
\end{proof}

\subsection{Propositions of Section~\ref{sec:bits}}
%
%
\begin{proposition}
  \label{prop:opt_loc_prf}
Let $C$ be a clause of $F$ and $F_c$ be the cluster of $F$ specified
by $C$.  Let $F'_c$ be a subset of $F_c$ that includes $C$ itself. Let
$F''_c=F_c \setminus F'_c$. Let $P$ be a set of certificate clauses
such that for every clause $B \in F'_c$ and for every primary literal
$l$ of $B$, the formula $P$ contains an \lpr{l} implied by $F
\setminus F''_c$.  Then $F\setminus F''_c$ (and hence $F$) is
unsatisfiable.
\end{proposition}
\begin{proof}
Denote by $F'$ the formula $F \setminus F''_c$.  Note that $C$ is in
$F'$ and $F'_c$ is the cluster of $F'$ specified by $C$. By our
assumption, for every clause $B$ of $F'_c$ and every literal $l$ of
$B$ shared with $C$, one can derive an \lpr{l} from $F'$. Hence \mprop
holds for $F'$. Then $F'$ (and so $F$) is unsatisfiable
\end{proof}

\section{Trivial SAT Procedure Based On \fm}
\label{app:str_fm}
%
%
\setlength{\intextsep}{2pt}
\begin{wrapfigure}{L}{2in}
\small
\begin{tabbing}
aa\=b\=cc\= dd\= \kill
\tfm\!\!$(\prob{X'}{F(X',X'')})$\\
\tb{\scriptsize{1}}\>  while ($\mi{true}$) \{\\
\tb{\scriptsize{2}}\Tt if ($\mi{unsat}(\Sub{F}{X''})$) return(\ti{False}) \\
\tb{\scriptsize{3}}\Tt   $v := \mi{PickVar}(X',F)$ \\
\tb{\scriptsize{4}}\Tt   if $(v = \mi{nil})$ return(\ti{True}) \\
\tb{\scriptsize{5}}\Tt   $X' = X' \cup \s{v^*}$ \\
\tb{\scriptsize{6}}\Tt   $F^* := \mi{Subst}(F,v,v^*)$ \\
\tb{\scriptsize{7}}\Tt   $H := \mi{PQE}(\prob{X'}{\mi{eq}(v,v^*) \wedge F^*})$\\
\tb{\scriptsize{8}}\Tt   $F := F^* \cup H$ \}\\
\end{tabbing} 
\vspace{-20pt}
\caption{SAT by FM}
\vspace{10pt}
\label{fig:str_fm}
\end{wrapfigure}

 In this appendix, we describe a simple SAT
procedure mentioned in Subsection~\ref{ssec:fm_compl} that is based
on \fm. We reuse the notation of Subsection~\ref{ssec:fm_compl}. That
is $F(X)$ is a formula to be checked for satisfiability $X$ is
partitioned into $X'$ and $X''$. The SAT procedure in question is
called \tfm. It accepts \prob{X'}{F(X',X'')} that is equisatisfiable
to $F$. It outputs \ti{True} (respectively \ti{False}) if $F$ is
satisfiable (respectively unsatisfiable). Recall that \Sub{F}{X''}
specifies the set of clauses of $F$ depending only on variables of
$X''$. We assume that $X''$ is much smaller than $X'$. So, checking
the satisfiability of
\Sub{F}{X''} is much easier than that of $F(X)$. Fig.~\ref{fig:str_fm}
shows the pseudocode of \tfm.

\tfm runs a \ti{while} loop. It starts by checking if the initial
formula \Sub{F}{X''} is unsatisfiable (line 2). If so, \tfm
returns \ti{False}. Otherwise, \tfm picks a variable $v \in X'$ for
melting $F$ (line 3). The only requirement for $v$ is that the current
formula $F$ contains at least one pair of clauses that can be resolved
on $v$. If no such variable is found, $F$ is satisfiable (line
4). Indeed, the formula \Sub{F}{X''} at this point is satisfiable.
Let \pnnt{x}{''} be a full assignment to $X''$
satisfying \Sub{F}{X''}.  Since $F$ does not have clauses resolvable
on a variable of $X'$, then \pnnt{x}{''} can be extended to an
assignment satisfying $F$ without producing any conflict.

If $v$ is found, \tfm performs melting of $F$ in $v$ (lines 5-8).
First, \tfm adds the twin variable $v^*$ to the set of quantified
variables $X'$ (line 5). Then \tfm produces a formula $F^*$ obtained
by replacing the variable $v$ with $v^*$ in (some) clauses of $F$
(line 6).  After that, \tfm builds a solution $H(X'')$ to the PQE
problem of taking $\mi{eq}(v,v^*)$ out of
\prob{X'}{\mi{eq}(v,v^*) \wedge F^*} (line 7). That is
\prob{X'}{\mi{eq} \wedge F^*} $\equiv H \wedge
\prob{X'}{F^*}$. (Recall that $\mi{eq}(v,v^*)$ specifies the
equivalence of $v$ and $v^*$.) In the final step of the loop, the
formula $F$ is replaced with $F^* \cup H$ and a new iteration begins.

\section{Efficient Computing Of Formulas $H_i$}
\label{app:fm_ec}
In this appendix, we describe a more efficient way to compute formulas
$H_1,\dots,H_k$ introduced in Section~\ref{sec:fm_for_ec}.  Let
$F^*_i$ denote the subset of clauses of $F^*$ specifying the gates
between $\mi{Cut}_i$ and the inputs of $M',M''$. (The definition of
$F^*$ is given in Section~\ref{sec:fm_for_ec}.) So, \smallskip
$F^*_{i-1} \subset F^*_i$. We will also assume that the last formula
$F^*_k$ includes the clauses of $\overline{\mi{eq}}(z',z'')$.  So
$F^*_k = F^*$. Let $W_i$ specify the variables of $F^*_i$ minus those
of $\mi{Cut}_i$. (Note that this definition of $W_i$ is different from
the one given in Section~\ref{sec:fm_for_ec}.) So, $W_{i-1} \subset
W_i$ and $W_k = W$.  Let $H_1$ be a formula obtained by taking
$\mi{Eq}(X,X^*)$ out of
\prob{W_1}{\mi{Eq} \wedge F^*_1}, \ie \prob{W_1}{\mi{Eq} \wedge F^*_1}
$\equiv H_1 \wedge \prob{W_1}{F^*_1}$. Let $H_i$, $i=2,\dots,k$, denote
the formula obtained by taking $H_{i-1}$ out of \prob{W_i}{H_{i-1}
  \wedge F^*_i}, \ie \prob{W_i}{H_{i-1} \wedge F^*_i} $\equiv H_i
\wedge \prob{W_i}{F^*_i}$. The \tb{semantics} of the sequence
$H_1,\dots,H_k$ is the same as in Section~\ref{sec:fm_for_ec}. Namely,
every next formula $H_i$ makes the previous formula $H_{i-1}$
\tb{redundant} (because $H_{i-1}$ is redundant in $H_i \wedge
\prob{W_i}{H_{i-1} \wedge F^*_i}$).

One can show that each formula $H_i$ is also a solution to the PQE
problem of taking $\mi{Eq}$ out of \prob{W_i}{\mi{Eq} \wedge F^*_i},
\ie \prob{W_i}{\mi{Eq} \wedge F^*_i} $\equiv H_i \wedge
\prob{W_i}{F^*_i}$, $i=1,\dots,k$. Since $F^*_k = F^*$ and $W_k = W$,
then \bm{H_k} \tb{specifies} the formula $H$ of
Proposition~\ref{prop:ec_by_pqe}.

One can boost the efficiency of computing $H_i$ from $H_{i-1}$ by
doing this \ti{approximately}. (To derive $H_k$ implying
$\mi{eq}(z',z'')$, it may suffice to use approximate $H_i$,
$i=1,\dots,k\!-\!1$.) This makes sense, since $F^*_{i-1} \subset
F^*_i$ and finding $H_i$ becomes harder as $i$ grows. An approximate
$H_i$ is obtained by using a \ti{subset} of $F^*_i$ when taking
$H_{i-1}$ out of \prob{W_i}{H_{i-1} \wedge F^*_i}.  (This subset has
to include at least the clauses specifying all gates between
$\mi{Cut}_{i-1}$ and $\mi{Cut}_i$.)  Approximation gives greater
efficiency at the expense of completeness. Now, \ec cannot guarantee
that $H$ implying $\mi{eq}(z',z'')$ is always produced if $M',M''$ are
equivalent.

\newpage
\section{Generation Of Local Proofs Used In Experiment}
\label{app:gen_loc_prf}
%
%
\setlength{\intextsep}{2pt}
\begin{wrapfigure}{l}{1.7in}
\small
\vspace{-10pt}
\begin{tabbing}
aaa\=bb\=cc\= dd\= \kill
\flp\!\!$(F)$\\
\tb{\scriptsize{1}}\> $P := \emptyset$ \\
\tb{\scriptsize{2}}\> $C := \mi{PickCls}(F)$  \\
\tb{\scriptsize{3}}\> $F_c := \mi{FormClust}(F,C)$   \\
\tb{\scriptsize{4}}\> for every $B \in F_c$  \\
\tb{\scriptsize{5}}\Tt for every $l \in \mi{Lits}(B)$ \{\\
\tb{\scriptsize{6}}\ttt if $(l \not\in \mi{Lits}(C))$ cont. \\
\tb{\scriptsize{7}}\ttt $B' := \mi{FlipLit}(B,l)$  \\
\tb{\scriptsize{8}}\ttt $P := P \cup \s{B'}$ \}\\
\tb{\scriptsize{9}}\> return($P$)\\
\end{tabbing} 
\vspace{-20pt}
\caption{\flp}
\label{fig:form_lp}
\end{wrapfigure}

In this appendix, we describe the procedure from
Section~\ref{sec:exper_lp} for generating local proofs. Given a
formula $F$, this procedure builds a set of clauses $P$ that forms a
local proof under the assumption $F \imp P$. We will call this
procedure \flp. The pseudocode of \flp is shown in
Fig.~\ref{fig:form_lp}. The complexity of \flp is $O(|F|\ast{k})$,
where $k$ is the length of the longest clause of $F$. This efficiency
is due to the fact that one does not need to prove $F \imp P$.

\flp starts with setting $P$ to an empty set (line 1). Then it picks a clause $C$
of $F$ (line 2) and forms the cluster $F_c$ specified by $C$ (line
3). After that, \flp runs two nested loops. Namely, for every clause
$B$ of the cluster $F_c$ (line 4) and for every literal $l$ of $B$
(line 5), \flp does the following. First, it checks if $l$ is a
primary literal, \ie whether $l$ is shared with $C$ (line 6). If not,
then this literal is skipped.  Otherwise, \flp forms a clause $B'$
that is an \lpr{l} for $B$, under the assumption $F \imp B'$ (line
7). $B'$ is obtained from $B$ by negating the literal $l$. Then $B'$
is added to $P$ (line 8).  When the outer loop is over, the set $P$
forms a local proof returned by \flp (line 9).

\stepcounter{example}
\ti{Example \theexample.}
Here we give an example of how \flp works.  Let $F = C_1 \wedge
C_2 \wedge C_3 \wedge \dots \wedge C_n$ where $C_1 = x_1 \vee x_2$,
$C_2 = x_1 \vee \overline{x}_3 \vee x_4$ and $C_3 = x_2 \vee
x_5 \vee \overline{x}_6$. Let $C_1,C_2,C_3$ be the only clauses of $F$
with positive literals of $x_1$ and $x_2$.

Assume that \flp picks the clause $C_1$ in line 2. Then it builds the
cluster $F_{c_1} = \s{C_1,C_2,C_3}$ in line 3.  Assume that $C_1$ is
the first clause of $F_{c_1}$ chosen in the outer loop.  Assume that
$x_1$ is the first literal of $C_1$ chosen in the inner
loop. Then \flp builds the clause $B'_1 = \overline{x}_1 \vee x_2$
obtained by flipping $x_1$ in $C_1$.  The clause $B'_1$ is
an \lpr{x_1} of $C_1$, assuming $F \imp B'_1$. The last
action of the first iteration of the inner loop is adding $B'_1$ to $P$.

In the second iteration of the inner loop, \flp builds $B''_1 =
x_1 \vee \overline{x}_2$, which is an \lpr{x_2} of $C_1$ under the
assumption $F \imp B''_1$. Then $B''_1$ is added to $P$.  At this
point, the inner loop terminates.

Assume in the second iteration of the outer loop, $C_2$ is picked. The
literals $\overline{x}_3$ and $x_4$ are skipped in the inner loop as
non-primary. When processing the primary literal $x_1$, the clause
$B_2 = \overline{x}_1 \vee \overline{x}_3 \vee x_4$ is
generated. $B_2$ is an \lpr{x_1} for $C_2$ under the assumption
$F \imp B_2$. The clause $B_2$ is added to $P$.

Finally, when the clause $C_3$ is picked in the outer loop, the clause
$B_3 = \overline{x}_2 \vee x_5 \vee \overline{x}_6$ is generated and
added to $P$. After that, \flp returns\linebreak $P = B'_1 \wedge
B''_1 \wedge B_2 \wedge B_3$. The latter gives a local proof under
the assumption $F \imp P$ $\blacksquare$

In Section~\ref{sec:exper_lp}, we used \flp to build local proofs for
random and BMC formulas. For a random formula $\mi{rand}_i$, we picked
a primary clause $C$ so that the size of the cluster $F_c$ was small.
However, in reality, one could pick \ti{any} clause of $F$ as a
primary clause $C$ when building a local proof $P$. That is, no matter
how $C$ was chosen, \kis had to do a lot of work to prove
$F\!\wedge\!P$ unsatisfiable.

For the BMC formulas, the choice of a primary clause was
different. Let $F$ be a formula specifying a combinational circuit $M$
obtained by unfolding a sequential circuit. A clause of $F$ was
selected as primary if it specified the functionality of a gate of $M$
that was fed directly by inputs of $M$. When the primary clause
specified a gate whose inputs were fed by outputs of other gates,
proving $F \wedge P$ unsatisfiable was easy.

\section{Verification Of Local Proof}
\label{app:lp_ver}

In this appendix, we describe the procedure \vlp mentioned in
Section~\ref{sec:exper_lp}.  Fig.~\ref{fig:ver_lp} shows the
pseudocode of \vlp.  \vlp accepts a formula $F$ and a set of
certificate clauses $P$. The assumption here is that $F \imp P$ holds.
\vlp returns either a subset $P'$ of $P$ that forms a local proof
or an empty set if it fails to find a local proof.

%
%
\setlength{\intextsep}{2pt}
\begin{wrapfigure}{l}{2in}
\small
\begin{tabbing}
aaa\=bb\=cc\=dd\=ee\=ff\=gg\=hh \kill
\vlp\!\!$(F,P)$\\
\tb{\scriptsize{1}}\> for every $C \in F$ \{ \\
\tb{\scriptsize{2}}\Tt $P':= \emptyset$ \\
\tb{\scriptsize{3}}\Tt  $F_c := FormClust(F,C)$ \\
\tb{\scriptsize{4}}\Tt  $\mi{clust\_fnd} := \mi{true}$ \\
\tb{\scriptsize{5}}\Tt for every $B \in F_c$ \{   \\
\tb{\scriptsize{6}}\ttt    for every $l \in \mi{Lits}(B)$ \{ \\
\tb{\scriptsize{7}}\tttt      if $(l \not\in \mi{Lits}(C))$ cont. \\
\tb{\scriptsize{8}}\tttt      $\mi{cert\_fnd} := \mi{false}$ \\
\tb{\scriptsize{9}}\tttt      for every $D \in P$ \\
\tb{\scriptsize{10}}\ttttt        if $(\mi{Certif}(D,B,l))$\{ \\
\tb{\scriptsize{11}}\tttttt          $\mi{cert\_fnd} := \mi{true}$ \\
\tb{\scriptsize{12}}\tttttt         $P' := P' \cup \s{D}$ \\
\tb{\scriptsize{13}}\tttttt          break \} \\
\tb{\scriptsize{14}}\tttt        if ($\mi{cert\_fnd}=\mi{false}$) \{ \\
\tb{\scriptsize{15}}\ttttt         $\mi{clust\_fnd} := \mi{false}$ \\
\tb{\scriptsize{16}}\ttttt          break \} \} \\
\tb{\scriptsize{17}}\ttt      if ($\mi{clust\_fnd} = \mi{false}$) \\
\tb{\scriptsize{18}}\tttt break \}\\
\tb{\scriptsize{19}}\Tt      if ($\mi{clust\_fnd}$) return($P'$)\} \\
\tb{\scriptsize{20}}\> return($\emptyset$)\\
\end{tabbing} 
\vspace{-20pt}
\caption{\vlp}
\label{fig:ver_lp}
\end{wrapfigure}

The complexity of \vlp is $O(|F|\ast|P|\ast{k}_1\ast{k_2})$ where
$k_1$ and $k_2$ are the maximal length of clauses of $F$ and $P$
respectively.  If $|P|$, $k_1$ and $k_2$ are small, \vlp runs in
near-linear time. (This efficiency is obviously due to the fact that
the hardest part of proof verification, namely, showing that $F \imp
P$, is assumed to be done.)

\vlp runs four nested loops. The outermost loop (lines 1-19) enumerates
clauses of $F$. Namely, for every clause $C \in F$, \vlp builds the
cluster $F_c$ and checks if it specifies a local proof $P'$. If so,
the outermost loop terminates (line 19).  Otherwise, the set $P'$ is
cleared and a new iteration of the outermost loop begins (line 2).The
loop of level 2 (lines 5-18) enumerates the clauses of $F_c$. If for
some clause $B \in F_c$ and for some primary literal $l$ of $B$, the
set $P$ does not have an \lpr{l}, this loop terminates reporting that
$F_c$ does not specify a local proof (line 18).

The loop of level 3 (lines 6-16) enumerates literals of the clause $B$
above. If, for some primary literal $l$ of $B$, the set $P$ does not
have an \lpr{l}, this loop terminates (line 16). Finally, the
innermost loop enumerates the clauses of $P$.  If some clause $D \in
P$ is an \lpr{l} for the literal $l$ of the clause $B$, then $D$ is
added to $P'$, and the loop of level 3 tries a new literal of $B$.

\section{Local Proofs And Blocked Clauses}
\label{app:blk_cls}
In this appendix, we relate local proofs and blocked clauses. In
Subsection~\ref{ssec:sim_proc}, we describe a procedure for simulating
a local proof by blocked clauses. In Subsection~\ref{ssec:sim_lp}, we
argue that although blocked clauses can simulate a local proof, this
is unlikely to be efficient in general. On the other hand, it is not
clear if blocked clauses can be efficiently simulated by local proofs
either. In Subsection~\ref{ssec:blkd_cls}, we substantiate some of our
claims by experiments.
%
%
\subsection{Procedure for simulating a local proof}
\label{ssec:sim_proc}
%
%
\setlength{\intextsep}{2pt}
\begin{wrapfigure}{L}{1.5in}
\small
\vspace{-15pt}
\begin{tabbing}
aa\=b\=cc\= dd\= \kill
\slp\!\!$(C,F_c,P)$\\
\tb{\scriptsize{1}}\> $\mi{Blk} := \emptyset$  \\
\tb{\scriptsize{2}}\> while (\ti{True}) \{\\
\tb{\scriptsize{3}}\Tt $\pnt{s}\!:=\!\mi{SAT}(F_c\!\wedge\!P\!\wedge\!\mi{Blk})$ \\
\tb{\scriptsize{4}}\Tt if $\pnt{s} \neq \mi{nil}$ \{\\
\tb{\scriptsize{5}}\ttt $B := BlkCls(\pnt{s},C,F_c)$  \\
\tb{\scriptsize{6}}\ttt $\mi{Blk} := \mi{Blk} \cup \s{B}$  \\
\tb{\scriptsize{7}}\ttt continue \} \\
\tb{\scriptsize{8}}\Tt  return \ti{Blk}\}  \\
\end{tabbing} 
\vspace{-20pt}
\caption{\slp}
\label{fig:lp_sim}
\end{wrapfigure}

 Fig.~\ref{fig:lp_sim} shows the pseudocode of \slp, a procedure for
simulating a local proof by blocked clauses. Let $F$ be a formula, $C$
be a clause of $F$, and $F_c$ be the cluster specified by $C$. Let $P$
be a local proof. Namely, \mprop holds for $F_c$ and $P$.

\slp accepts $F_c$ and $P$. It starts by setting the set of blocked
clauses \ti{Blk} to an empty set (line 1). Then \slp runs a \ti{while}
loop. It begins by checking if $F_c\wedge\!P\!\wedge\!\mi{Blk}$ is
satisfiable (line 3). If so, a satisfying assignment \pnt{s} is
produced. Then \slp generates a clause $B$ blocked at a variable of
$C$ and falsified by \pnt{s} (line 5). (Since $F_c$ contains all the
clauses of $F$ sharing a literal with $C$, a clause blocked in $F_c$
is \tb{also blocked} in $F$.) After that, $B$ is added to \ti{Blk} and
a new iteration begins.  If $F_c \wedge P \wedge \mi{Blk}$ is
unsatisfiable, the current set \ti{Blk} \tb{simulates} the local proof
$P$. That is, one can prove $F$ unsatisfiable using only the machinery
of blocked clauses and the facts that a) $F_c$ is the cluster
specified by $C$ and b) $F \imp P$. So, \slp returns \ti{Blk} (line
8).

%
%
\setlength{\intextsep}{2pt}
\begin{wrapfigure}{L}{1.4in}
\small
\begin{tabbing}
aa\=b\=cc\= dd\= \kill
\bcl\!\!$(\pnt{s},C,F_c)$\\
\tb{\scriptsize{1}}\> $l_1 := \mi{SatLit}_1(\pnt{s},C)$  \\
\tb{\scriptsize{2}}\> $F'_c := \mi{FormCls}(F_c,l_1)$\\
\tb{\scriptsize{3}}\> $B := \{\,\overline{l_1}\,\}$ \\
\tb{\scriptsize{4}}\> for every $C'\in F'_c$ \{  \\
\tb{\scriptsize{5}}\Tt  $l_2 := \mi{SatLit}_2(\pnt{s},C',l_1)$ \\
\tb{\scriptsize{6}}\Tt  $B = B \cup \{\,\overline{l_2}\,\}$\}\\
\tb{\scriptsize{7}}\>  $B^* := \mi{DisjLits}(B)$ \\
\tb{\scriptsize{8}}\> return($B^*$)\\
\end{tabbing} 
\vspace{-20pt}
\caption{\bcl}
\label{fig:blk_cls}
\vspace{5pt}
\end{wrapfigure}

Fig.~\ref{fig:blk_cls} shows the pseudocode of \bcl called by \slp to
generate a blocked clause.  \bcl accepts an assignment \pnt{s}
satisfying $F_c \wedge P \wedge \mi{Blk}$, the primary clause $C$ and
the cluster $F_c$ specified by $C$. It returns a blocked clause
falsified by \pnt{s}.

\bcl starts by picking a literal $l_1$ of  $C$ satisfied by \pnt{s}
(line 1). Then it forms the subset $F'_c$ of $F_c$ consisting of the
clauses sharing the literal $l_1$ (line 2).  (Every clause of $F_c$
shares at least one literal with $C$ but not necessarily the literal
$l_1$.) Then \bcl initializes the set $B$ of literals from which the
final blocked clause is formed with the negation of $l_1$ (line 3).

After that, \bcl runs a \ti{for} loop where it enumerates the clauses
of $F'_c$. Let $C'$ be a clause of $F'_c$.  First, \bcl picks a second
literal $l_2$ of $C'$ satisfied by \pnt{s} (line 5). (Since, $P$ is a
local proof, \mprop holds. This entails that \pnt{s} satisfies \ti{at
least} two literals of every clause of $F'_c$. Indeed, for every
clause $C'$ of $F'_c$ and every literal $l$ shared with $C$, the set
$P$ contains an \lpr{l} that excludes any assignment satisfying only
literal $l$ of $C'$.) Then the negation of $l_2$ is added to the set
$B$ (line 6).

After the loop is over, \bcl returns the clause $B^*$ obtained by
disjoining the literals of $B$. By construction, \pnt{s} falsifies
$B^*$. Besides, $B^*$ is blocked in $F_c$ at the variable
corresponding to $l_1$. As mentioned above, $B^*$ is also blocked in
$F$.

%
%
\subsection{Relating resolution, local proofs and blocked clauses}
\label{ssec:sim_lp}
We will refer to resolution enhanced by \ti{local proofs} and
\ti{blocked clauses} as \bm{\mi{Res}\hspace{-0.5pt}^\mi{lp}} and
\bm{\mi{Res}\hspace{-0.5pt}^\mi{bc}} respectively. We will refer to \ti{pure}
resolution as \bm{\mi{Res}\hspace{-0.5pt}^\mi{pu}}.

\begin{wraptable}{l}{2.8in}
\small
\caption{\ti{Hard random formulas.\\ Using blocked clauses}}
\label{tbl:rnd_form_blk}
\vspace{-10pt}
\setlength\extrarowheight{2pt}
\scriptsize
\begin{center}
  \begin{tabular}{|c|c|c|c|c|c|c|c|} \hline
    name & \multicolumn{2}{c|}{cluster}& \multicolumn{5}{c|}{$\mi{Blk}$ are blocked clauses, $F'\!=\!F\!\wedge\!P\!\wedge\!\mi{Blk}$}   \\ \cline{2-8}
    &$|F_c|$&$|P|$& \multicolumn{3}{c|}{$\mi{Blk}$ is gen. by \slp}  &  \multicolumn{2}{c|}{$\mi{Blk}$ is random}  \\\cline{4-8}
    &&& $|\mi{Blk}|$&gener. &$F'$,kis-& $|\mi{Blk}|$  &$F'$,kis- \\
    & &&            &time &sat time & &sat time  \\ \hline
    $\mi{rand}_1$ & 9 & 11 & 10 &0.0s & 0.0s & 100 &1.3s \\ \hline
    $\mi{rand}_2$ & 10 & 12 & 32 & 0.0s & 0.0s &200 & 12s    \\ \hline
    $\mi{rand}_3$ & 15 & 17 & 676 &0.3s & 0.0s &2,000 & 1m\,12s\\ \hline
    $\mi{rand}_4$ & 20 & 22 & 96 & 0.0s & 0.0s & 500 & 11m \\ \hline
    $\mi{rand}_5$ & 12 & 14 & 48 &0.0s & 0.0s & 200 & 29m \\ \hline
    $\mi{rand}_6$ & 17 & 19 & 72  &0.0s & 0.0s & 300  & 3h\,36m   \\ \hline
    $\mi{rand}_7$ & 14 & 16 &2,052 & 4.1s & 0.1s & 4,000 & 42h \\ \hline    
  \end{tabular}
\end{center}
\end{wraptable}

\rbc is known to be  exponentially more powerful than
\rpu~\cite{gen_ext_resol}. The fact that simulating \rlp naturally involves
blocked clauses implies that \rlp is likely more powerful than \rpu as
well.  As we saw above, \rbc can simulate \rlp. However, the number of
blocked clauses the procedure \slp generates is \ti{exponential} in
the cluster size. This means that the simulation provided by \slp is,
in general, not efficient. On the other hand, there is no trivial way
to simulate adding a blocked clause by local proofs (a local proof is
equivalent to adding \ti{multiple} blocked clauses).  So, one can
\ti{speculate} that neither \rbc nor \rlp  simulates the other efficiently.

\subsection{Some experiments}
\label{ssec:blkd_cls}
\begin{wraptable}{l}{2.8in}
  \small
 \vspace{3pt}
\caption{\ti{BMC formulas.\\ Using blocked clauses}}
\label{tbl:pc_form_blk}
\vspace{-10pt}
\setlength\extrarowheight{2pt}
\scriptsize
\begin{center}
  \begin{tabular}{|c|c|c|c|c|c|c|c|} \hline
    name & \multicolumn{2}{c|}{cluster} & \multicolumn{5}{c|}{$\mi{Blk}$ are blocked clauses, $F'\!=\!F\!\wedge\!P\!\wedge\!\mi{Blk}$ }   \\ \cline{2-8}
    &$|F_c|$ &$|P|$ &\multicolumn{3}{c|}{$\mi{Blk}$ is gen. by \slp}  &  \multicolumn{2}{c|}{$\mi{Blk}$ is random} \\\cline{4-8}
    & & & $|\mi{Blk}|$&gener. &$F'$, kis-& $|\mi{Blk}|$  &$F'$, kis-   \\
  &  & &    &time & sat time & & sat time  \\ \hline
6s101 &5&7 &  3   & 0.0s & 5.1s  &  30  & 2m 37s   \\ \hline
6s154 &69& 71& 50,603 &   $24$h*  &  0.5s  & 60,000 &  8m      \\ \hline
6s165 &4& 6&  3 &  0.0s  &  0.0s  & 30    &  2m\,29s  \\ \hline
6s168 &72&74&45,880 &  $24$h* & 0.1s   & 50,000   &   2m     \\ \hline
6s284 &5&7& 3    & 0.0s   &   18s    &  30   & 25m      \\ \hline
6s343 & 5   & 7 & 3 & 0.0s &  1.7s & 30 &  5m\,17s      \\ \hline
6s372 &5 & 7&  3   &0.0s& 2.3s&    30 &   29m     \\ \hline
  \end{tabular}
\end{center}
\end{wraptable}

In this subsection, we provide some experimental results using the
formulas we employed in Section~\ref{sec:exper_lp}. Let $F$ be a
formula to solve by SAT, $F_c$ be the cluster specified by a clause
$C \in F$ and $P$ be a local proof related to $F_c$. In
Section~\ref{sec:exper_lp}, we showed that \kis, a high-quality CDCL
SAT solver, needed a lot of time to prove $F \wedge P$ unsatisfiable
(\ie to find a \ti{resolution} proof rather than the local proof $P$).
Here we check what happens if one adds to $F \wedge P$ a set \ti{Blk}
of blocked clauses. We used two methods to form \ti{Blk}: by the
procedure \slp above and by generating blocked clauses randomly.

Table~\ref{tbl:rnd_form_blk} and~\ref{tbl:pc_form_blk} show the
results. Both tables have the same structure. The first column gives
the name of the formula.  The next two columns describe the size
$|F_c|$ of a cluster and the size $|P|$ of the local proof generated
for $F_c$. The next three columns show the results when \slp
generated \ti{Blk}.  The first column of the triple gives the number
of clauses in \ti{Blk}.  The next column shows \slp's runtime and the
last column gives \kis's runtime when solving $F \wedge
P \wedge \mi{Blk}$. The last two columns provide the results
when \ti{Blk} was generated randomly. The first of these two columns
gives the number of clauses in \ti{Blk} and the second column
shows \kis's runtime when solving $F \wedge P \wedge \mi{Blk}$.

One can make the following conclusions. First, blocked clauses carry
significant power. In many cases, adding a (relatively) small number
of blocked clauses dramatically reduced the run time of \kis. For
instance, for the random formula $\mi{rand}_7$, adding \ti{Blk}
reduced the run time from 41 hours (when running \kis on $F \wedge P$)
to 4.1 seconds. Note however, that this is a \ti{non-deterministic}
power of blocked clauses since they were generated by \slp that served
as an ``\ti{oracle}''. When we generated even more clauses
\ti{randomly}, the run time on $F \wedge P \wedge \mi{Blk}$ did
not change much compared with $F \wedge P$. Second, generation of
blocked clauses even by \slp \tb{becomes problematic} for large
clusters.  For instance, for BMC formulas 6s154 and 6s168, \slp could
not finish in 24 hours and was terminated after generating
$5.1\!\ast\!10^4$ and $4.6\!\ast\!10^4$ blocked clauses, respectively.

\section{Generation Of \ti{LP}-clauses}
\label{app:lp_cls}
%
%
%

In Subsection~\ref{ssec:lp_cls} we introduced the notion of
an \lc. This appendix gives an example of how an \lc is generated.  It
also provides a proof of correctness of the generation procedure.

\begin{example}
\label{exmp:lp_cls}
Let $F$ be a formula, and $C_1 = x_1 \vee x_2 \vee x_3 \vee x_4$ be a
clause of $F$. Assume $C_1$ is not satisfied by the current assignment
\pnt{q} explored by \elp. Assume, for the sake of simplicity, that
\mprop holds for the unoptimized cluster $F_{c_1}$ specified by $C_1$ and the current set
of certificates $P$ in subspace \pnt{q} (\ie \mprop holds for the
cluster \cof{(F_{c_1})}{q} and the set \cof{P}{q}). Let $Q$ specify
the \lc corresponding to the local proof specified by $P$. The clause
$Q$ consists of a subset of literals falsified by \pnt{q}. Informally,
$Q$ includes the literals that are relevant to making \cof{P}{q} a
local proof.  (An \lc for an \ti{optimized} cluster is built
similarly. The difference is that only \ti{some} clauses of $F_{c_1}$
are relevant.)

\ti{First}, $Q$ includes the primary literals (\ie those of $C_1$)
falsified by \pnt{q}. Assume, for instance, that \pnt{q} contains
assignments $x_1=0$ and $x_3=0$. Then $Q$ includes the literals $x_1$
and $x_3$. \ti{Second}, for every clause $C \in F_{c_1}$ that is
satisfied by \pnt{q}, the \lc $Q$ contains the negation of a literal
of $C$ that is satisfied by \pnt{q}. Suppose $C = x_4 \vee x_9$
and \pnt{q} assigns $x_9=1$. Then $Q$ contains the literal
$\overline{x}_9$. \ti{Third}, for every clause $B \in P$ that serves as a
certificate for a clause $C \in F_{c_1}$, the \lc $Q$ contains the
literals of $B$ falsifying which makes \cof{B}{q} imply \cof{C}{q}.
Assume, for instance, that $B=\overline{x}_5 \vee x_7 \vee x_8$, $C =
x_2 \vee \overline{x}_5$, and \pnt{q} contains assignments $x_7=0$ and
$x_8=0$. So, $B$ is an \lpr{x_2} for $C$ in subspace \pnt{q} because
\cof{B}{q} = $\overline{x}_5$ and \cof{C}{q} = $x_2 \vee \overline{x}_5$.
Then $Q$ contains the literals $x_7$ and $x_8$ $\blacksquare$
\end{example}

Propositions~\ref{prop:lp_subsp} and~\ref{prop:lp_cls} below apply to
generation of \lcs. For the sake of simplicity, we assume that the
local proof in question is formulated for the full (\ie unoptimized)
cluster specified by a clause. The proof of correctness of the
generation procedure for an \ti{optimized} cluster looks similar.
%
%
%
\begin{proposition}
\label{prop:lp_subsp}
Let $C$ be a clause of a formula $F(X)$ and \pnt{q} be an assignment
to $X$ that does not satisfy/falsify $C$.  Let $F_c$ denote the
cluster of $F$ specified by $C$. Let $P$ be a set of certificates.
Suppose that for every clause $B \in F_c$ that is not satisfied by
\pnt{q} and for every literal $l$ of $B$ shared with $C$ and
unassigned by \pnt{q}, the set $P$ contains a clause that is
an \lpr{l} for $B$ in subspace \pnt{q}. Then $F$ is unsatisfiable in
subspace
\pnt{q}.
\end{proposition}
\begin{proof}
The satisfiability of $F$ in subspace \pnt{q} is equivalent to the
satisfiability of \cof{F}{q}.  (Recall that \cof{F}{q} denotes the
formula obtained from $F$ by removing the clauses satisfied by \pnt{q}
and removing the literals falsified by \pnt{q} from the remaining
clauses.)  Note that the cluster of \cof{F}{q} specified by \cof{C}{q}
and the certificates of \cof{P}{q} satisfy the conditions of
\mprop. So, \cof{F}{q} is unsatisfiable.  Hence, $F$ is unsatisfiable
in subspace \pnt{q}.
\end{proof}
%
%
\begin{proposition}
\label{prop:lp_cls}
Assume the conditions of Proposition~\ref{prop:lp_subsp} hold for a
formula $F$, a set of certificates $P$, a clause $C \in F$ and an
assignment \pnt{q}.  Assume $F_c$ denotes the cluster of $F$ specified
by $C$. Then every clause Q satisfying the three conditions
below is falsified by \pnt{q} and implied by $F$.  First, Q
includes every literal of $C$ falsified by \pnt{q}.  Second, for every
clause $B$ of $F_c$ satisfied by \pnt{q}, Q contains the
negation of a literal of $B$ satisfied by \pnt{q}. Third, for every
clause $B$ of $F_c$ unsatisfied by \pnt{q} and for every literal $l$
of $B$ shared with $C$ and unassigned by \pnt{q}, Q contains
literals of a clause $D \in P$ that is an \lpr{l} for $B$ in subspace
\pnt{q}. Namely, Q contains all the literals of $D$ but
those of $C$ or their negations.
\end{proposition}
\begin{proof}
The assignment \pnt{q} falsifies Q because the latter consists only of
literals falsified by \pnt{q}. Let \pnt{r} denote the shortest
assignment falsifying the clause Q. By construction, \pnt{r} is a
subset of \pnt{q}. To prove that Q is implied by $F$, it suffices to
show that $F$ is unsatisfiable in the subspace \pnt{r}. Note that $F$
satisfies the conditions of Proposition~\ref{prop:lp_subsp} in
subspace \pnt{r}. First, $C$ is not satisfied/falsified
by \pnt{r}. Second, for every clause of $B$ of $F_c$ that is not
satisfied by \pnt{r} and for every literal $l$ of $B$ shared with $C$
and unassigned by \pnt{r}, there is a clause of $P$ that is an \lpr{l}
for $B$ in subspace \pnt{r}.  Hence, $F$ is unsatisfiable in
subspace \pnt{r}.
\end{proof}


\end{document}